\documentclass[twocolumn]{revtex4-1}
\usepackage{graphicx}
\usepackage{dcolumn}
\usepackage{bm}
\usepackage{amsmath}
\usepackage{physics}

\newcommand{\tens}[1]{%
	\mathbin{\mathop{\otimes}\limits_{#1}}%
}

\let\oldAA\AA
\renewcommand{\AA}{\text{\normalfont\oldAA}}

\usepackage{amsmath}
\usepackage{subfigure}
\usepackage{hyperref}
\usepackage[normalem]{ulem}
\hypersetup{
    colorlinks,
    citecolor=blue,
    filecolor=black,
    linkcolor=blue,
    urlcolor=blue
}
\usepackage{epstopdf}
 \usepackage{relsize}

\newcommand*{\rom}[1]{\expandafter\@slowromancap\romannumeral #1@}
\title{Report}

\setcitestyle{square}

\begin{document}


\title{A realistic non-local heat engine based on Coulomb coupled systems }

\author{Aniket Singha}
\affiliation{%
Department of Electronics and Electrical Communication Engineering,\\
Indian Institute of Technology Kharagpur, Kharagpur-721302, India\\
}%





\begin{abstract}


Optimal non-local heat-engines,  based on Coulomb-coupled systems, demand a sharp step-like change  in the energy resolved system-to-reservoir coupling around the ground state of quantum-dots \cite{coulomb_TE1,coulomb_TE2,coulomb_TE3,coulomb_TE4,coulomb_TE6,coulomb_TE7}. Such  a sharp step-like transition in the system-to-reservoir coupling  cannot be achieved in a realistic scenario. Here, I propose realistic design for non-local heat engine based on Coulomb-coupled system, which circumvents the  need for any change in the   system-to-reservoir coupling, demanded by the optimal set-ups discussed in literature. I demonstrate that  an intentionally introduced  asymmetry (or energy difference)  in the  ground state configuration between adjacent tunnel coupled quantum dots, in conjugation with Coulomb coupling, is sufficient  to convert the stochastic fluctuations from a non-local heat source into a directed flow of  thermoelectric current.   The performance, along with the regime of operation, of the  proposed heat engine is then theoretically investigated using quantum-master-equation (QME) approach. It is demonstrated that the theoretical maximum power output for the proposed set-up is limited to about $50\%$ of the optimal design. Despite a lower performance compared to the optimal set-up, the novelty of the proposed design lies in the conjunction of fabrication simplicity along with reasonable power output. At the end, the sequential transport processes leading to a performance deterioration of the proposed set-up are analyzed and a method to alleviate such transport processes is discussed. The set-up proposed in this paper    can be used to design and  fabricate high-performance  non-local cryogenic heat engines.

\end{abstract}
\maketitle
\section{Introduction}
\begin{figure}
	\includegraphics[width=.4\textwidth]{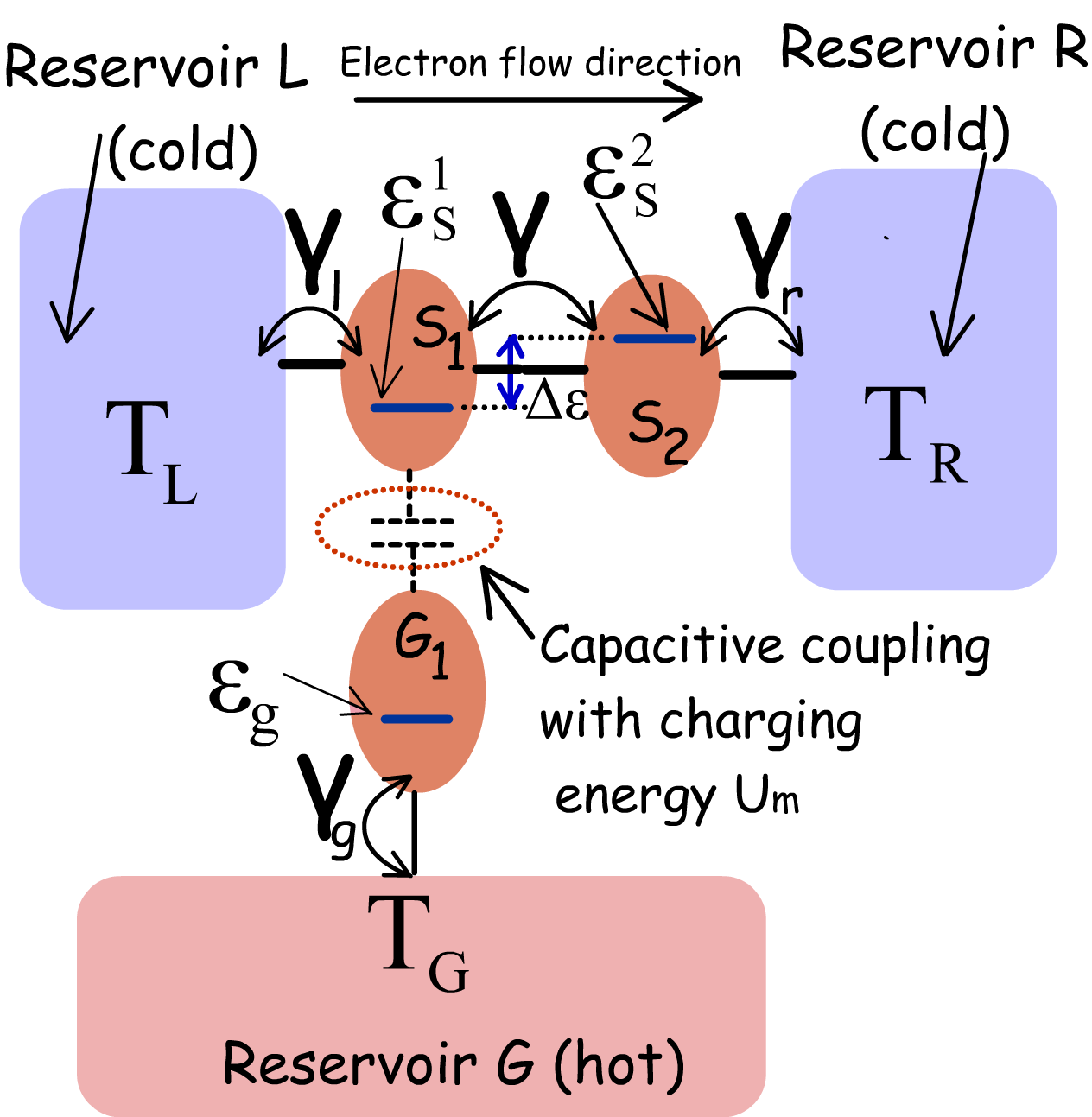}
	\caption{Schematic diagram of the proposed non-local heat-engine based on Coulomb-coupled quantum dots. The entire system consists of three dots $S_1,~S_2$ and $G_1$ which are electrically coupled to the reservoirs $L,~R$ and $G$ respectively. The dots $S_1$ and $S_2$ are tunnel coupled, while $S_1$ and $G_1$ are Coulomb coupled. The ground states of $S_1$ and $S_2$ form a staircase configuration with $\varepsilon_s^2 = \varepsilon_s^1+\Delta \varepsilon$. In the proposed arrangement, current can be driven between the cold reservoirs $L$ and $R$ by absorbing thermal energy from the hot reservoir $G$.}
	\label{fig:Fig_1}
\end{figure}
With the progress in fabrication and scaling technology, efficient heat harvesting in lower dimensional systems has gained a lot of attention \cite{aniket,aniket_heat1,aniket_heat2,nanoinclusion1,nanoinclusion2,nanoinclusion3,theory1,akshay,kim1,kim2,theory2,theory3,aniket_cool1,aniket_cool2}. One of the major issues affecting heat harvesting performance in nano-systems is the drastic lattice heat flux resulting from a small spatial separation between the heat reservoir and the heat sink. The large lattice heat flux severely limits the overall efficiency of the heat engine and poses a major performance issue in cases where supply of heat energy is limited. Tailoring the lattice thermal conductance, in an attempt to gain enhanced harvesting efficiency  generally affects the current path, thereby deteriorating the peak harvested power. As such, one of the crucial focus of the  modern thermoelectric community is to facilitate an independent optimization of  the  electron transport path and lattice heat  conduction path, by introducing a spatial separation between the current path and the heat source  \cite{sep1,sep2,sep3}.   This phenomenon of harvesting heat from a reservoir, which is spatially separated from the current conduction path, is known as non-local heat harvesting \cite{sep1,sep2,sep3,coulomb_TE1,coulomb_TE2,coulomb_TE3,coulomb_TE4,coulomb_TE5,coulomb_TE6}. Thus, non-local heat engines are three terminal systems where power is delivered between two terminals while extracting heat from the third terminal which is spatially separated from the path of current flow. \color{black} In this case,  tailoring the  lattice heat transport path, in an attempt to  gain enhanced efficiency, can be accomplished without altering the current conduction path.  Recently designs and concepts of non-local heat engines and refrigerators using Coulomb  coupled quantum dots have been proposed and  explored theoretically \cite{coulomb_TE1,coulomb_TE2,coulomb_TE3,coulomb_TE4,coulomb_TE5,coulomb_TE6} as well as experimentally \cite{nl_expmt_1} in literature.  However, the   operation of such non-local heat engines demand a sharp step-like change in the system-to-reservoir coupling around the ground state energy \cite{coulomb_TE1,coulomb_TE2,coulomb_TE3,coulomb_TE4,coulomb_TE5,coulomb_TE6}, which is impossible to achieve in a practical scenario \cite{nl_expmt_1}. In this paper, I propose a realistic design strategy to accomplish non-local heat harvesting using Coulomb coupled quantum dots. Unlike the optimal non-local heat-engine based on Coulomb coupled systems \cite{coulomb_TE1,coulomb_TE2,coulomb_TE3,coulomb_TE4,coulomb_TE5,coulomb_TE6}, the proposed design doesn't demand a change in the system-to-reservoir coupling near the ground state.  The performance  proposed heat-engine is then evaluated and compared with the optimal set-up. It is demonstrated that the performance of the proposed heat engine hovers around $50\%$ of the optimal set-up. However, the novelty of the proposed set-up is the conjugation of fabrication simplicity along with a reasonable power output. At the end, the processes leading to a performance deterioration of the proposed set-up are discussed and analyzed. This paper is organized as follows. In Sec.~\ref{design}, I illustrate the proposed design strategy and elaborate the transport formulation employed to analyze the thermoelectric performance of the same. Next, Sec \ref{results} elaborates a detailed analysis on the heat harvesting performance and regime of operation of the proposed heat-engine. A performance comparison between the proposed heat engine and the optimal set-up is also conducted along with a brief discussion on the transport processes leading to a performance deterioration of the proposed heat engine. I conclude this paper briefly in Sec. \ref{conclusion}. The Appendix Sec. elaborates some intuitive and conceptual understanding of the proposed heat engine operation, as well as details the derivations of the equations employed to investigate the performance of the heat engine.\\
\section{Proposed design and transport formulation}\label{design}
\indent The proposed heat engine, schematically demonstrated in Fig.~\ref{fig:Fig_1}, consists of three dots $S_1,~S_2$ and $G_1$ which  are electrically coupled to the reservoirs $L$, $R$ and $G$ respectively.  I will now discuss the ground state configuration and other features of the system consisting of the three dots $S_1,~S_2$ and $G_1$. $S_1$ and $S_2$ are tunnel coupled to each other, while $G_1$ is capacitively coupled to $S_1$. The ground states of $S_1$ and $S_2$ form a stair-case configuration with $\varepsilon_s^2= \varepsilon_s^1+\Delta \varepsilon$.   Any electronic tunneling between the dots  $S_1$ and $G_1$ is suppressed via suitable fabrication techniques. Energy exchange between the two dots is,  however, possible via Coulomb coupling \cite{cap_coup_1,cap_coup_2,cap_coup3,cap_coup_4,cap_coup_5}. In the optimal Coulomb-coupled system based heat-engine discussed in literature, an asymmetric system-to-reservoir coupling is required for heat harvesting \cite{coulomb_TE6,coulomb_TE7}. In the proposed set-up, instead of the asymmetric system-to-reservoir coupling, the system itself is made asymmetric with respect to the reservoir $L$ and $R$ by choosing an energy difference between the ground states of  the adjacent quantum dots $S_1$ and $S_2$, with $\varepsilon_s^2=\varepsilon_{s}^1+\Delta \varepsilon$.  The purpose of such arrangement is to deliver power to an external load connected between $L$ and $R$ while extracting heat from reservoir $G$.  Another equivalent realistic set-up, based on Coulomb coupled systems, that can be employed for efficient non-local heat harvesting is demonstrated in Fig.~\ref{fig:app_2} and discussed briefly in Appendix \ref{app_e}. \color{black}\color{black} Due to random electron exchange between the dot $G_1$ and the reservoir $G$, the total energy of the system  fluctuates stochastically with time.  For a Coulomb coupled system with constant energy resolved system-to-reservoir coupling and with symmetric ground state configuration with respect to reservoir $L$ and $R$, the stochastic fluctuations result in the same probability of electron transfer from $L$ to $R$ and from $R$ to $L$, making the average current zero, although there is a non-zero component of electronic heat flow between $G$ and $L(R)$ (See Appendix \ref{app_a}). However, for a Coulomb coupled system with asymmetric ground state configuration, the same fluctuation can be converted to a directed current flow when $T_G\neq T_{L(R)}$ (Appendix \ref{app_c}).   I will demonstrate via numerical calculations and theoretical arguments (given in Appendix~\ref{app_c}) that in the system detailed above, thermoelectric power can be delivered between the terminals $L$ and $R$ (with a net electronic flow from $L$ to $R$) by extracting heat energy from the reservoir $G$ when $T_G>T_L(R)$. Thus, power can be delivered between terminals non-local to the heat source.  The excess energy   $\Delta \varepsilon=\varepsilon_s^2-\varepsilon_s^1$, required for the electrons to tunnel from $S_1$ to $S_2$ is supplied from the reservoir $G$ via Coulomb coupling.  Coming to the realistic fabrication possibility of such a system, due to the recent advancement in solid-state nano-fabrication technology, triple and quad  quantum dot systems with and without Coulomb coupling have already been realized experimentally \cite{mqd1,mqd2,mqd3,mqd4,mqd5,mqd6}. In addition, it has been experimentally demonstrated that quantum dots that are far from each other in space, may be bridged to obtain strong Coulomb coupling, along with  excellent thermal isolation between the hot and cold reservoirs \cite{cap_coup_1,cap_coup_2,cap_coup3,cap_coup_4,cap_coup_5}. Also, the bridge may be fabricated between two specific quantum dots to drastically enhance their mutual Coulomb coupling, without affecting the electrostatic energy of the other quantum dots \cite{cap_coup_1,cap_coup_2,cap_coup3,cap_coup_4,cap_coup_5}. \color{black} Thus, the change in electron number $n_{S_1}~(n_{G_1})$ of the dot $S_1$ ($G_1$)  influences the electrostatic energy of the  dot $G_1$ ($S_1$).
   In general,  the total increase in electrostatic energy $U$ of the configuration  demonstrated in Fig.~\ref{fig:Fig_1} (a), consisting  of three dots, due to fluctuation in electron number  can be given by (Refer to Appendix \ref{app_b} for detailed derivation): 
   \begin{widetext}
 	\begin{equation}
 		U(n_{S_{1}},n_{G_{1}},n_{S_{2}})=\sum_{x }U^{self}_{x}\left(n_{x}^{tot}-n_x^{eq}\right)^2   +\sum_{(x_{1},x_{2})}^{x_1 \neq x_2} U^m_{x_1,x_2}\left(n_{x_1}^{tot}-n_{x_1}^{eq}\right)\left(n_{x_2}^{tot}-n_{x_2}^{eq}\right) 
 	\end{equation}
 	\end{widetext}
where $n_x^{tot}$  is the total  electron number, and $U^{self}_x=\frac{q^2}{C^{self}_{x}}$ is the electrostatic energy due to self-capacitance $C^{self}_{x}$ (with the surrounding leads) of    quantum dot `$x$' (Refer to Appendix \ref{app_b} for details). $U^m_{x_1,x_2}$ is the electrostatic energy arising out of  Coulomb coupling between two different quantum dots that are separated in space (Appendix \ref{app_b}) and $n_x^{eq}$ is the total number of electrons present in dot $x$ in equilibrium at $0K$ and is determined by the minimum possible electrostatic energy of the system. $n_x=n_x^{tot}-n_x^{eq}$ is the total number of excess electrons added in the ground state of the dot $x$ due to stochastic fluctuations from the reservoirs  (Refer to Appendix \ref{app_b} for details). Here, a minimal physics based model is used to investigate the heat engine performance under the assumption that the change in potential  due self-capacitance is much greater than than the average thermal voltage $kT/q$ or the applied bias voltage $V$, that is $U^{self}_x=\frac{q^2}{C^{self}_{x}}>> (kT,~qV)$. Hence, electron occupation probability or transfer rate via the Coulomb blocked energy level, due to self-capacitance, is negligibly small.  The analysis  of the entire system of dots may hence be approximated by limiting the maximum number of electrons in each  dot to one. Thus the analysis of the entire system may be limited to eight multi-electron  levels, which I denote by the electron occupation number in the ground state of each quantum dot. Hence, a possible state of interest in the system may be denoted  as $\ket{n_{S_1},n_{G_1},n_{S_2}}=\ket{n_{S_1}}\tens{} \ket{n_{G_1}} \tens{} \ket{n_{S_2}}$, where $n_{S_1},n_{G_1},n_{S_2}\in (0,1)$,  are used to denote the number of electrons present in the ground-states of $S_1,~G_1$ and $S_2$ respectively. I also assume that the electrostatic coupling between $S_1,~S_2$ and between $S_2,~G_1$  are negligible, such that, for all practical purposes under consideration, $U^m_{S_1,S_2} \approx 0$ and $U^m_{G_1,S_2} \approx 0$. Since, the electronic transport and ground states in $S_1$ and $G_1$ are mutually coupled, I  treat the pair of dots $S_1$ and $G_1$ as a  sub-system ($\varsigma_1$), $S_2$ being the complementary sub-system ($\varsigma_2$) of the entire system consisting of three dots \cite{sispad}.  The state probability of   $\varsigma_1$ is denoted by $P_{i,j}^{\varsigma_1}$,  $i$ and $j$ being the number of electrons in the ground state of dot $S_1$ and $G_1$ respectively. $P_k^{\varsigma_2}$, on the other hand, denotes the probability of occupancy of the dot $S_2$ in the sub-system $\varsigma_2$. It can be shown that if $\Delta \varepsilon$ is much greater than the ground state broadening due to system-to-reservoir   coupling, then the interdot tunneling rate between $S_1$ and $S_2$ is optimized when $\varepsilon_s^1+U^m_{S_1,G_1}=\varepsilon_s^2$, that is when $\Delta \varepsilon=U^m_{S_1,G_1}$ \cite{sispad} (See Appendix \ref{app_b}). To evaluate the optimal performance of the proposed heat-engine, I hence assume $\Delta \varepsilon=U^m_{S_1,G_1}$ \cite{sispad}. Henceforth, I would simply represent $U^m_{S_1,G_1}$ as $U_m$.  Under the  assumption stated above, the  equations governing sub-system state probabilities in steady state can be derived  as \cite{sispad} (See Appendix \ref{app_b} for detailed derivation):	
\small
\begin{widetext}
	\begin{align}
	&  -P_{0,0}^{\varsigma_1}\{f_L(\varepsilon_s^1)+f_G(\varepsilon_g)\}+P_{0,1}^{\varsigma_1}\{1-f_G(\varepsilon_g)\}+P_{1,0}^{\varsigma_1}\{1-f_L(\varepsilon_s^1)\}=0 \nonumber \\
	& -P_{1,0}^{\varsigma_1}\left\{1-f_L(\varepsilon_{s}^1)+f_G(\varepsilon_g+U_m)\right\}+P_{1,1}^{\varsigma_1}\left\{1-f_G(\varepsilon_g+U_m)\right\}+P_{0,0}^{\varsigma_1}f_L(\varepsilon_s^1) \nonumber \\
	&-P_{0,1}^{\varsigma_1}\left\{1-f_g(\varepsilon_{g}^1)+f_L(\varepsilon_s^1+U_m)+\frac{\gamma}{\gamma_c}P^{\varsigma_2}_1\right\}+P_{0,0}^{\varsigma_1}f_G(\varepsilon_g)+P_{1,1}^{\varsigma_1}\left\{1-f_L(\varepsilon_s^1+U_m)+\frac{\gamma}{\gamma_c}P^{\varsigma_2}_{0}\right\} = 0 \nonumber \\
	& -P_{1,1}^{\varsigma_1}\left\{[1-f_g(\varepsilon_{g}^1+U_m)]+[1-f_L(\varepsilon_s^1+U_m)]+\frac{\gamma}{\gamma_C}P^{\varsigma_2}_0\right\}+P_{1,0}^{\varsigma_1}f_G(\varepsilon_g+U_m) +P_{0,1}^{\varsigma_1}\left\{f_L(\varepsilon_s^1+U_m)+\frac{\gamma}{\gamma_c}P^{\varsigma_2}_{1}\right\}=0
	\label{eq:first_sys}
	\end{align} 
\end{widetext}
\begin{align}
& -P_{0}^{\varsigma_2}\{f_R(\varepsilon_s^2)+\frac{\gamma}{\gamma_c}P_{1,1}^{\varsigma_1}\}+P_1^{\varsigma_2}\left\{1-f_R(\varepsilon_{s}^2)+\frac{\gamma}{\gamma_c}P^{\varsigma_1}_{0,1}\right\}=0\nonumber \\
&-P_1^{\varsigma_2}\{1-f_R(\varepsilon_{s}^2)+\frac{\gamma}{\gamma_c}P^{\varsigma_1}_{0,1}\}+P_{0}^{\varsigma_2}\left\{f_R(\varepsilon_s^2)+\frac{\gamma}{\gamma_c}P_{1,1}^{\varsigma_1}\right\}=0,
\label{eq:second_sys}
\end{align}
\normalsize where $\gamma_c=\gamma_l(\varepsilon)=\gamma_r(\varepsilon)=\gamma_g(\varepsilon)$ and $\gamma$ are related to the reservoir-to-system tunnel coupling and the inter-dot tunnel coupling respectively \cite{sispad,dattabook}, $\varepsilon$ being the independent energy variable.  In the above set of equations, $f_{\lambda}(\varepsilon)$ denotes the probability of occupancy of the reservoir $\lambda$ at energy $\varepsilon$. For the purpose of  calculations in this paper, I assume an equilibrium Fermi-Dirac statistics at the reservoirs. Hence, $f_{\lambda}(\varepsilon)$ is given by:
\begin{equation}
f_{\lambda}(\varepsilon)=\left(1+exp\left\{\frac{\varepsilon-\mu_{\lambda}}{kT_{\lambda}}\right\}\right)^{-1},
\end{equation} 
where $\mu_{\lambda}$ and $T_{\lambda}$ respectively denote the quasi-Fermi energy and temperature of the reservoir $\lambda$. From the set of Eqns.~\eqref{eq:first_sys} and \eqref{eq:second_sys}, it is clear that an electron in $S_1$ can tunnel into $S_2$ only when the ground state in the dot $G_1$ is occupied with an electron. The set of  Eqns.~\eqref{eq:first_sys} and \eqref{eq:second_sys} are coupled to each other and may be solved using any iterative method. Here, I use Newton-Raphson iterative method to solve the steady-state values of sub-system probabilities. On calculation of the sub-system state probabilities $P_{i,j}^{\varsigma_1}$ and $P_k^{\varsigma_2}$, the electron current flow into (out of) the system from the reservoirs $L ( R)$ can be given as:
\small
\begin{figure}[!htb]
	\includegraphics[width=.4\textwidth]{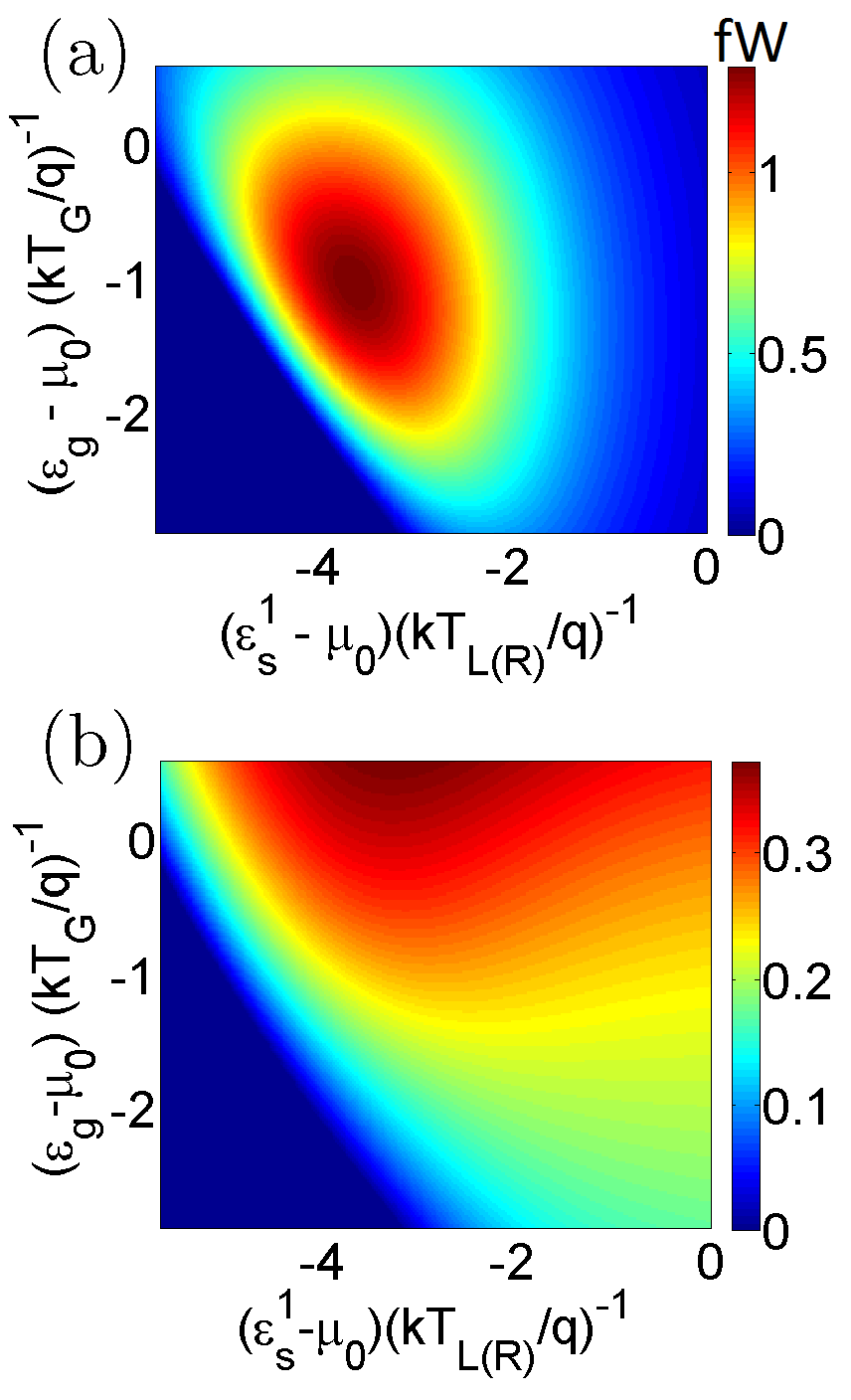}
	\caption{Variation of the proposed heat engine performance with variation in the the ground states $\varepsilon_g$ and $\varepsilon_s^1$ for $U_m=3.9meV~(\approx 6\frac{kT}{q})$ and  $V=1.3meV~(\approx 2\frac{kT}{q})$. Colour plot demonstrating the variation in (a) generated power ($P$) and (b) efficiency $\eta/\eta_c$. $T=\frac{T_{L(R)}+T_G}{2}=7.5K$ is the average temperature between the heat source and the heat sink. The efficiency of generation is measured with respect to the Carnot efficiency $\eta_c=1-T_{L(R)}/T_G$  }
	\label{fig:Fig_2}
\end{figure}
\begin{align}
I_L= & q\gamma_c \times \left\{P^{\varsigma_1}_{0,0}f_L(\varepsilon_s^1)+P^{\varsigma_1}_{0,1}f_L(\varepsilon_s^1+U_m)\right\} \nonumber \\ &- q\gamma_c P^{\varsigma_1}_{1,0}\{1-f_L(\varepsilon_s^1)\}- q\gamma_c P^{\varsigma_s^1}_{1,1}\{1-f_L(\varepsilon_s^1+U_m)\} \nonumber \\
I_R= & -q\gamma_c \times \left\{P^{\varsigma_2}_{0}f_R(\varepsilon_s^1)-P^{\varsigma_2}_{1}\{1-f_R(\varepsilon_s^1)\}\right\} ,
\label{eq:final}
\end{align}
\normalsize
In addition, the electronic component of heat flow from the  reservoir $G$ can be given by:
\small
\begin{equation}
I_{Qe}=U_m \gamma_c\left\{P^{\varsigma_1}_{10}f_G(\varepsilon_g+U_m)-P^{\varsigma_1}_{11}\{1-f_G(\varepsilon_g+U_m)\}\right\}
\label{eq:heat}
\end{equation}  
\normalsize
Interestingly, we note that Eqn.~\eqref{eq:heat} is not directly dependent on  $\varepsilon_g$. This is due to the fact that the net electronic current into or out of the reservoir $G$ is zero (See Appendix \ref{app_b} for details). Next, I use the set of Eqns.~\eqref{eq:first_sys}, \eqref{eq:second_sys}, \eqref{eq:final} and \eqref{eq:heat}, to evaluate the thermoelectric generation performance of the the set-up demonstrated in  Fig.~\ref{fig:Fig_1}. To analyze the performance of the heat engine, I use a voltage-controlled set-up demonstrated in literature \cite{v1,v2,v3}, where a bias voltage $V$ is applied between $L$ and $R$, with the positive and negative terminals being connected to $L$ and $R$ respectively, to emulate the voltage drop due to thermoelectric current flow across an external load. Thus the bias voltage $V$ mimics the voltage drop between the terminals across which power is to be delivered.   Assuming the equilibrium electrochemical potential  across the entire set-up is $\mu_0$, and a voltage drop $V$ across the external load distributed symmetrically between the left and right system-to-reservoir interfaces, the quasi-Fermi levels at the reservoirs $L$, $R$ and $G$ may be written as $\mu_{L}=\mu_0- \frac{V}{2},~\mu_{R}=\mu_0+ \frac{V}{2}$ and $\mu_G=\mu_0$ respectively. The generated power ($P$) and efficiency ($\eta$) can be defined as:
\begin{align}
P&=I_{L(R)} \times V,\nonumber\\
\eta&=\frac{P}{I_Q},
\label{eq:pow_eff}
\end{align}
where $V$ is the applied potential bias in the voltage controlled model and $I_Q$ is the sum of electronic and lattice heat flux from the heat source ($G$). In the non-local heat engine the lattice heat flux can be favourably engineered \cite{supressk1,supressk2,nanowire_heat1,nanowire_heat2,nanoflake_heat} without affecting the current conduction path. In addition the lattice heat flux is generally independent of the the electronic heat conductivity and doesn't change with the system energy configuration \cite{phonon1,phonon2,phonon3,phonon4,phonon5,phonon6,phonon7,superlattice1}. Hence, to simplify the calculation,  I assume ideal condition by neglecting the lattice heat flux, as done in recent literature \cite{coulomb_TE1,coulomb_TE4,coulomb_TE5,coulomb_TE6,aniket,aniket_heat2,whitney,whitney2}. The generation efficiency for our case, can hence be defined as: 
\begin{equation}
\eta=\frac{P}{I_{Q_e}},
\label{eq:eff_act}
\end{equation}
where  $I_{Qe}$ is the electronic component of the heat current extracted from the heat source $G$ (defined in Eq.~\ref{eq:heat}).
 It should be noted that the efficiency given by Eq.~\eqref{eq:eff_act} is the maximum achievable efficiency under ideal conditions and is limited by thermodynamic considerations of heat to work conversion for the given system energy configuration. 
 \color{black} \\
To understand the operation of the proposed heat engine, let us assume that the system is initially in the  vacuum state $\ket{0,0,0}$, where the ground state of all the dots are empty. Now, let us consider a complete cycle $F \Rightarrow \ket{0,0,0}\rightarrow \ket{1,0,0}\rightarrow \ket{1,1,0}\rightarrow \ket{0,1,1}\rightarrow \ket{0,1,0}\rightarrow \ket{0,0,0}$. In this cycle, the system starts from the initial vacuum state $\ket{0,0,0}$. Next an electron enters into $S_1$ from $L$, with an energy $\varepsilon_{s}^1$, followed by an electron tunneling into $G_1$ from $G$ with an energy $\varepsilon_{g}+U_m$. Next, the electron in $S_1$ tunnels into $S_2$ with an energy $\varepsilon_s^2=\varepsilon_{s}^1+U_m$ and finally tunnels out of $S_2$ into reservoir $R$. The system, at last returns to its initial state when the electron in $G_1$, finally tunnels out into $G$ with an energy $\varepsilon_{g}$. In this cycle, an electron is transferred from reservoir $L$ to $R$ while absorbing a heat packet $U_m$ from $G$. In the reverse cycle, corresponding to the cycle $F$ described above, that is  $R \Rightarrow \ket{0,0,0}\rightarrow \ket{0,1,0}\rightarrow \ket{0,1,1}\rightarrow \ket{1,1,0}\rightarrow \ket{1,0,0}\rightarrow \ket{0,0,0}$, an electron is transferred from reservoir $R$ to $L$, while dumping a heat packet $U_m$ into reservoir $G$. It can be shown that when $T_G>T_{L(R)}$, under short-circuited condition between $L$ and $R$,  the rate of the forward cycle $F$ is greater than the reverse cycle $R$  (See  Appendix~\ref{app_b}), resulting in a net unidirectional flow of electrons from $L$ to $R$ while absorbing heat from reservoir $G$. Also for all other equivalent cycles,  resulting in transfer of electrons between $L$ and $R$ under short-circuited condition, it can be shown that a rate of forward sequence that results in electron transfer from $L$ to $R$ while absorbing heat from $G$ is greater than the reverse sequence when $T_G>T_{L(R)}$. Thus for $T_G>T_{L(R)}$, a net thermoelectric electronic flow occurs from $L$ to $R$ by extracting heat from the spatially separated reservoir $G$ and hence the set-up proposed in this paper constitutes a non-local heat engine.     Without loss of generality, I assume that $\gamma_c= 10^{-5}\frac{q}{h}$ and $ \gamma= 10^{-4}\frac{q}{h}$ (The effect of the ratio $\gamma/\gamma_c$ on the heat harvesting performance of the set-up is discussed in Appendix \ref{app_f}).  Such values of $\gamma$ and $\gamma_c$ limits electronic transport phenomena in the weak coupling regime, where the effects of cotunneling can be neglected and the transport through the entire system can be described in terms of the quantum master equation (QME) approach, given by the set of Eqns.~\eqref{eq:first_sys}-\eqref{eq:second_sys}. This simplifies the analysis to a great extent and facilitates an intuitive understanding of the computed results. \color{black}  The temperature of the reservoirs $L(R)$ and $G$ are assumed to be $T_{L(R)}=5K$ and $T_G=10K$. The average temperature between the hot and the cold reservoirs is, hence, given by $T=\frac{T_{L(R)}+T_G}{2}=7.5K$.\\
\begin{figure}[!htb]
	\includegraphics[width=.4\textwidth]{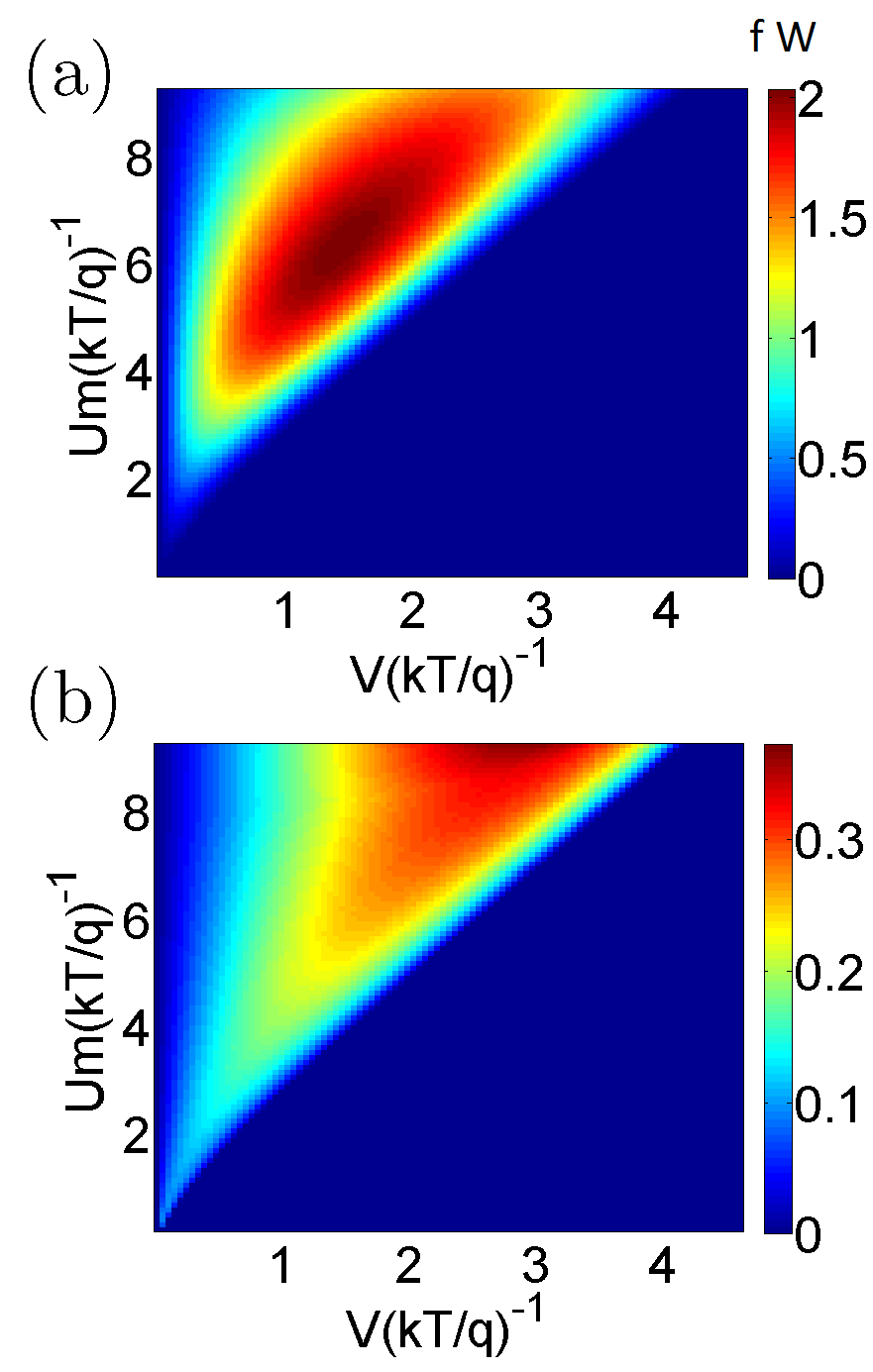}
	\caption{ Variation in the peak performance of the heat engine with variation in  Coulomb coupling energy $U_m$ and applied bias $V$. Colour plot depicting the (a) peak generated power $P_M$ and (b) efficiency at the peak generated power for a range of values of $V$ and $U_m$.   To find out the maximum power $P_M$ for a given value of $V$ and $U_m$, the ground states of the dots are tuned to optimal position.   $T=\frac{T_{L(R)}+T_G}{2}=7.5K$ is the average temperature between the heat source and the heat sink. The efficiency of maximum power ($P_M)$ generation is normalized with respect to the Carnot efficiency $\eta_c=1-T_{L(R)}/T_G$ }
	\label{fig:Fig_3}
\end{figure}
\section{Results}\label{results}
\begin{figure*}[!htb]
	\includegraphics[width=1.05\textwidth]{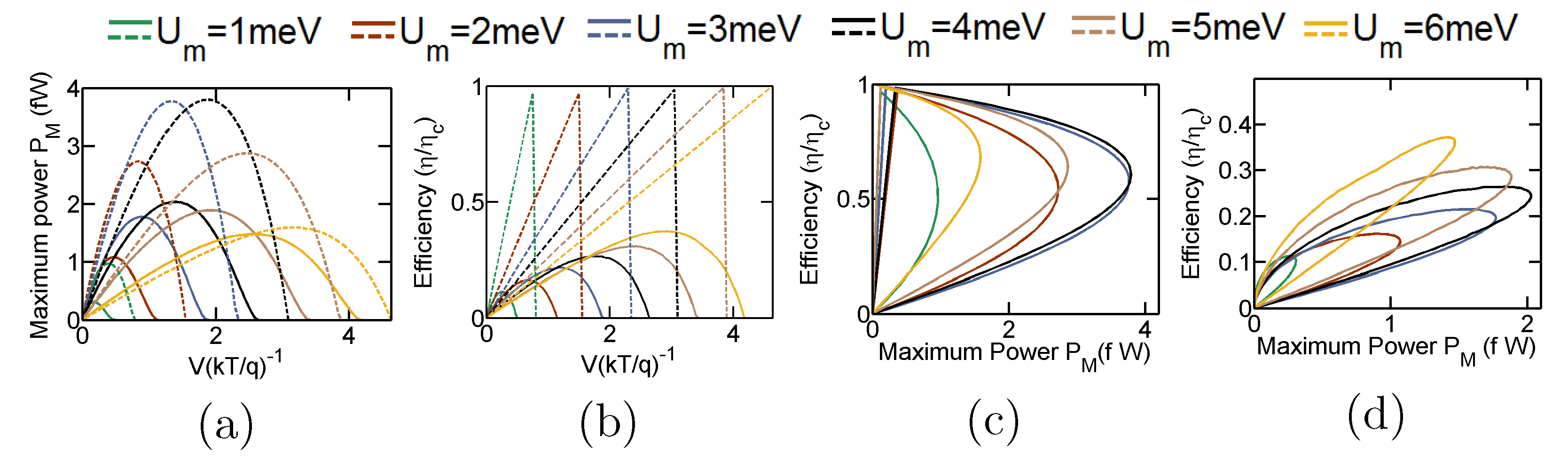}
	\caption{Performance comparison of the proposed non-local heat engine (solid lines) with the optimal set-up \cite{coulomb_TE1,coulomb_TE2,coulomb_TE3,coulomb_TE4} (dashed lines) for different values of the Coulomb coupling energy. Plot of (a) maximum generated power $P_M$ vs bias voltage $V$, (b) efficiency (with respect to Carnot efficiency) at the maximum generated power vs bias voltage, (c) maximum power $P_M$ vs efficiency at the maximum power for the optimal set-up (d) maximum power $P_M$ vs efficiency at the maximum power for the proposed design.   }
	\label{fig:Fig_4}
\end{figure*}
\indent In this section, I discuss the optimal operation regimes of the proposed heat engine. In addition, I conduct a performance comparison of the proposed heat engine with the optimal set-up discussed in literature and investigate the transport processes leading to a performance deterioration of the proposed set-up.  Fig.~\ref{fig:Fig_2} demonstrates the performance of the heat engine, in particular the generated power $P$ and efficiency of generated power ($\eta/\eta_c$) over a range of the positions of the ground states $\varepsilon_g$ and $\varepsilon_s^1$ for  $U_m=3.9meV~(\approx 6\frac{kT}{q})$ and  $V=1.3meV~(\approx 2\frac{kT}{q})$. It can be noted that the regime of heat harvesting corresponds to $\varepsilon_s^1$ lying a few $kT_{L(R)}$ below $\mu_0$. Such a behaviour can be expected since net interdot electron flow is optimized when $\varepsilon_s^1$ lies  a few $kT_{L(R)}$  below the equilibrium Fermi-energy.  This is because for $T_G>T_{L(R)}$, electrons must tunnel into $S_1$ with energy $\varepsilon_{s}^1$ and out of $S_2$ with energy $\varepsilon_{s}^2=\varepsilon_{s}
^1+U_m$ for finite power generation, which demands finite value of $f_L(\varepsilon_{s}^1)$ and $1-f_R(\varepsilon_s^2)$.  Both $f_L(\varepsilon_{s}^1)$ and $1-f_R(\varepsilon_s^2)$ are finite and large only  when $\varepsilon_s^1$ lies a few $kT_{L(R)}$ below $\mu_0$. In fact the product  $f_L(\varepsilon_{s}^1)\{1-f_R(\varepsilon_s^2)\}$ is maximized when $\varepsilon_{s}^1 \approx \mu_0-\frac{U_m}{2}$ (for small bias $V$). Since, $U_m\approx 6\frac{kT}{q}=8\frac{kT_{L(R)}}{2}$, the generated power is maximized when $\varepsilon_{s}^1\approx \mu_0-\frac{U_m}{2}=\mu_0-4\frac{kT_{L(R)}}{q}$, as demonstrated in Fig.~\ref{fig:Fig_2}. \color{black} Similarly, it can also be noted that for optimal heat harvesting, the ground state $\varepsilon_{g}$ must lie below a few $kT_G$ of the equilibrium Fermi energy $\mu_0$.   This can be understood by the following: if $\varepsilon_{g}-\mu_0<- few~kT_G$, then the ground state $\varepsilon_g$ is always occupied with an electron and so the asymmetry of the system ground states with respect to the reservoir $L$ and $R$ disappears. Hence a directional thermoelectric current flow is not possible \cite{coulomb_TE6} by rectifying the stochastic fluctuations of the heat source. On the other hand, when $\varepsilon_{g}-\mu_0> few~kT_G$, the probability of an electron tunneling into the reservoir $G_1$ with an energy $\varepsilon_g+U_m$ (provided that the ground state of $S_1$ is occupied) is negligibly small, resulting in the deterioration of the unidirectional current flow and power generation.  In other words, for the generation of finite thermoelectric power with $T_G>T_{L(R)}$, electrons must tunnel in and out of $G_1$ with energy $\varepsilon_{g}+U_m$ and $\varepsilon_{g}$ respectively for the absorption of heat energy from $G$, which calls for a finite value of both  $f_G(\varepsilon_{g}+U_m)$ and $1-f_G(\varepsilon_{g})$. Both these quantities are  finite and large only if $\varepsilon_{g}$ lies below a few $kT_G$ of the equilibrium Fermi energy $\mu_0$. In fact, the product $f_G(\varepsilon_{g}+U_m)\{1-f_G(\varepsilon_{g})\}$, and hence the generated power, is maximized when $\varepsilon_{g}\approx \mu_0-\frac{U_m}{2}$. Since, in this case, $U_m=6\frac{kT}{q}=4\frac{kT_G}{q}$, the generated power is maximized, when $\varepsilon_g \approx \mu_0-2\frac{kT}{q}$, as depicted in Fig.~\ref{fig:Fig_2}(a). \color{black}  Fig.~\ref{fig:Fig_2}(b) demonstrates the heat engine efficiency as a function of the ground state energy levels. The efficiency of heat harvesting decreases monotonically with decrease in $\varepsilon_g-\mu_0$. 
 This  is because, as $\varepsilon_{g}$ decreases, the probability occupancy of the ground state of $G_1$ increases, which increases the probability of reverse electron flux due to the  bias voltage. An increase in the reverse electron flux decreases the net thermoelectric current (and hence power) and thus degrades the efficiency. It should be noted that an equivalent trend of decrease in efficiency  can be found in bulk and lower dimensional thermoelectric engines as the equilibrium  Fermi-energy gradually moves inside the energy-bands \cite{aniket,aniket_heat2,akshay,whitney,whitney2}. The variation in generation efficiency with $\varepsilon_s^1$ is non-monotonic.  Initially as $\varepsilon_s^1$ decreases and moves away the Fermi energy (or electron transport window), the generated power increases (as discussed above) leading to an increase in efficiency. As  $\varepsilon_s^1$  further decreases, such that $\varepsilon_{s}^1+U_m$ enters the electron transport window,  the probability of reverse electronic flow due to voltage bias  increases, leading to a decrease in the total net thermoelectric current (discussed later). A decrease in the net thermoelectric current  leads to a deterioration in the generated power and hence, efficiency. \\ \color{black}
\indent The variation of the optimal performance of the heat engine with variation in the Coulomb coupling energy $U_m$ and applied bias $V$ is demonstrated in Fig.~\ref{fig:Fig_3}. In particular, Fig.~\ref{fig:Fig_3} (a) demonstrates the maximum generated power ($P_M$), while Fig.~\ref{fig:Fig_3} (b) demonstrates the efficiency at the maximum generated power for a range of values of the applied bias $V$ and the Coulomb coupling energy $U_m$.   To calculate the the maximum generated power $P_M$ for a given value of $V$ and $U_m$, the ground states of the dots are tuned to the optimal energy position. It should be noted that the maximum generated power is low for low values of $U_m$. This is due to the fact that the ground state of the dots approach symmetrical arrangement with respect to the reservoir $L$ and $R$ as $U_m$ approaches towards zero. Hence, the directional flow of electrons decreases. As $U_m$ increases, the asymmetry of the system increases resulting in an increase in directional electron flow, and hence, the maximum generated power \cite{coulomb_TE6}. With further increase in $U_m$, the maximum generated power reaches its peak and then decreases due to lower probability of an electron tunneling into $G_1$ with an energy $\varepsilon_g+U_m$, when the ground state of $S_1$ is already occupied.  Mathematically, the non-monotonic change in maximum generated power with $U_m$   can be explained in terms of the Eqns.~\eqref{eq:LR}-\eqref{eq:ratio_assy} developed in Appendix \ref{app_c} (Refer to Appendix \ref{app_c} for details). \color{black} For a fixed value of $U_m$, the maximum generated power first increases and then decreases with an increase in the bias voltage $V$. Such a behaviour is indeed expected from heat engines as the regime of operation approaches from short-circuited (zero voltage drop across the external load) condition to the open-circuited condition (zero net current across the external load) \cite{v3,whitney,whitney2}. Fig.~\ref{fig:Fig_3}(b) demonstrates the efficiency at the maximum generated power with variation in applied bias $V$ and Coulomb coupling energy $U_m$. The efficiency varies non-monotonically with the applied bias $V$, that is,the efficiency increases with an increase in $V$ as the regime of operation approaches the point of maximum power and then gradually decreases as the generated power decreases with the regime of operation approaching the open circuited condition. Such a trend can also be noted in bulk and lower dimensional heat engines \cite{aniket,aniket_heat2}. On the other hand  with an increase in $U_m$, the efficiency at the maximum power increases monotonically as $\varepsilon_{g}+U_m$ gradually surpasses the Fermi energy. An equivalent trend, again, can be noted in bulk and lower dimensional thermoelectric engines as the band-edge gradually  surpasses the Fermi-energy  \cite{aniket,v1,v2,v3,v5}.    Fig. \ref{fig:Fig_4} demonstrates a performance comparison of the proposed heat engine with the optimal non-local heat engine put forward in literature \cite{coulomb_TE1,coulomb_TE2,coulomb_TE3,coulomb_TE4}. The optimal non-local heat engine, discussed in literature, consists of a pair of Coulomb coupled quantum dots (demonstrated in Appendix Fig.~\ref{fig:app_1}.a) with asymmetric system-to-reservoir coupling \cite{coulomb_TE1,coulomb_TE2,coulomb_TE3,coulomb_TE4} given by $\gamma_l(\varepsilon)=\gamma_c \theta(\varepsilon_s^1+\delta \epsilon-\varepsilon),~\gamma_l(\varepsilon)=\gamma_c \theta(\varepsilon-\delta \epsilon-\varepsilon_s^1)$ and $\gamma_g(\varepsilon)=\gamma_c$, where $\delta \varepsilon <U_m$ and $\theta $ is Heaviside step function. \color{black} In particular, Fig.~\ref{fig:Fig_4}(a) and (b) demonstrate the variation in the maximum power $P_M$ and efficiency at the  maximum power respectively with applied load bias $V$ for different values of $U_m$ for both the proposed set-up (solid lines) and the optimal set-up (dashed lines). We note that the overall maximum generated power for the proposed design $P_{MAX}^{prop}$ is approximately $3.23 fW$, which is  about $50\%$ of the overall maximum power output of $6.04 fW$  for the optimal design $P_{MAX}^{opt}$. In both these cases the maximum power is generated around $U_m\approx 4meV(\approx 6\frac{kT}{q})$. As already discussed in literature \cite{coulomb_TE6}, the generation efficiency  for the optimal set-up increases linearly with the applied bias for a given value of $U_m$. The efficiency at the overall maximum power for our proposed design and the optimal set-up are $24.5\%$ and $60\%$   of the Carnot efficiency respectively. From Fig.~\ref{fig:Fig_4}(a) and (b), we also note that the open-circuit voltage (finite voltage at which out-put power just becomes zero)  for both the proposed design and the optimal set-up increases with an increase in $U_m$. In addition, it can also be noted that the open circuit voltage for the optimal set-up is slightly higher compared to the proposed design.  Fig. \ref{fig:Fig_4} (c) and (d)  demonstrate the the maximum power vs efficiency trade-off trend \cite{aniket} for the optimal design and the proposed design respectively for various values of $U_m$. We note that the power-efficiency trade-off for the optimal set-up is  mild compared to the proposed design.\\
\begin{figure}[!thb]
	\includegraphics[width=.5\textwidth]{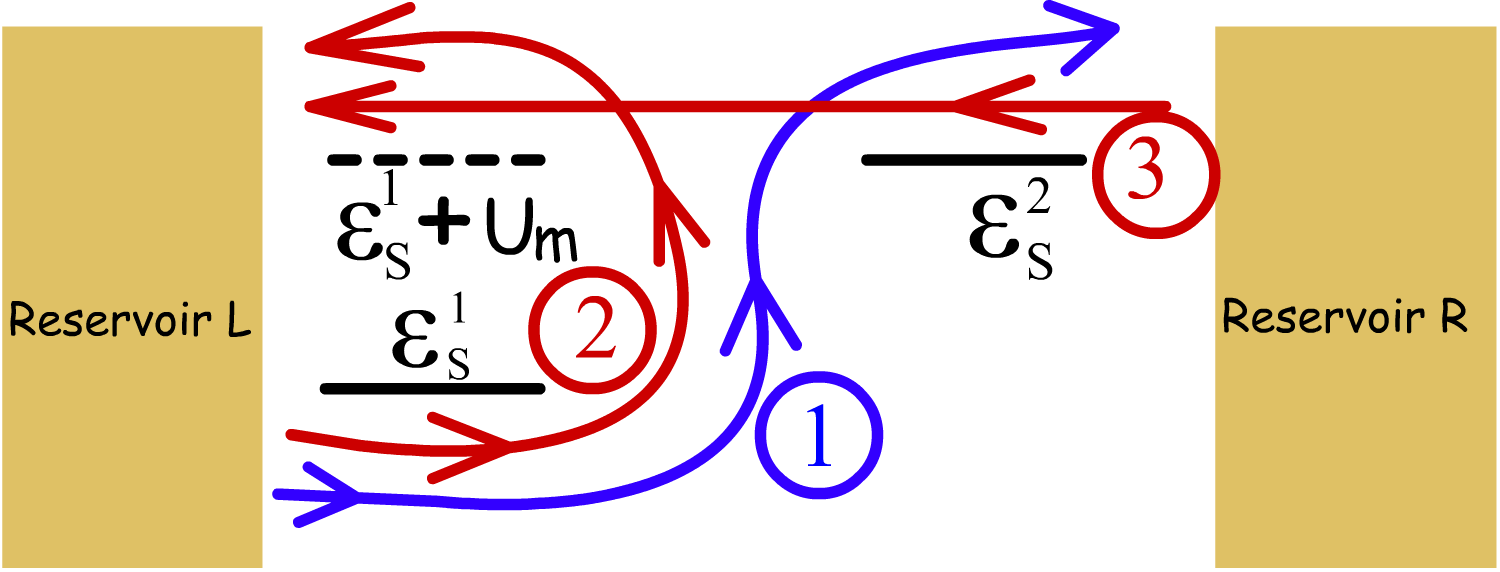}
	\caption{Schematic diagram demonstrating the components of electron flow between the reservoir $L$ and the system through the energy level $\varepsilon_{s}^1$ and the Coulomb blockaded level $\varepsilon_s^1+U_m$. Three current components are shown in the figure. (1) Electrons flow from the reservoir $L$ to  $R$ while absorbing a heat packet $U_m$ (per electron) from $G$. (2) Electrons enter into $\varepsilon_s^1$ from $L$. Next these electrons tunnel out of the system into $L$ through $\varepsilon_s^1+U_m$.  (3) These electrons flow in the direction of the voltage bias  (opposite to the thermoelectric current) and thus reduce the net current through the system. }
	\label{fig:Fig_6}
\end{figure}
\begin{figure}[!thb]
	\includegraphics[width=.39\textwidth]{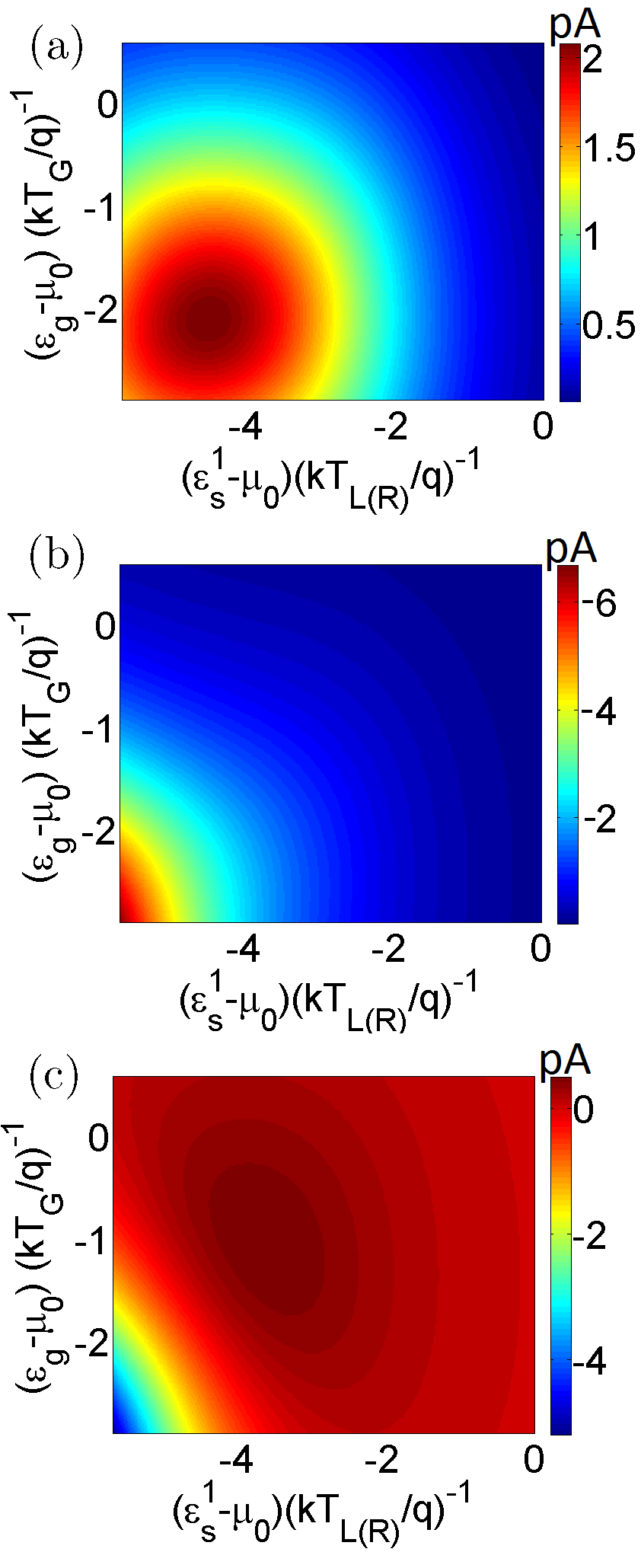}
	\caption{Colour plots demonstrating the  electron flow into the system from the reservoir $L$ with variation in the the ground states $\varepsilon_g$ and $\varepsilon_s^1$, for $U_m=3.9meV~(\approx 6\frac{kT}{q})$ and  $V=1.3meV~(\approx 2\frac{kT}{q})$, when the (a) ground state of the dot $G_1$ is not occupied (b) ground state of the dot is occupied (c) total average current between the system and the reservoir $L$. Interestingly,  when the ground state of the dot $G_1$ is occupied, the net electronic flow is directed from the system into the reservoir $L$. Since, the net direction of current for thermoelectric generation should be from reservoir $L$ to $R$, the effect of net electronic flow from the system into the reservoir $L$ is to reduce the generated power and efficiency.}
	\label{fig:Fig_5}
\end{figure}
\indent I end the discussion with a brief description of the processes leading to a deterioration in generated power and efficiency for the proposed set-up.  First, let us consider the  cycles leading to electron transport from the reservoir $L$ to the reservoir $R$ against the applied bias while absorbing a heat packet $U_m$ from reservoir $G$.  Consider the cycle $\ket{0,0,0}\rightarrow \ket{1,0,0})\rightarrow\ket{1,1,0}\rightarrow\ket{0,1,1}\rightarrow\ket{0,1,0}\rightarrow\ket{0,0,0}$. In this cycle, the system starts with an initial vacuum state $\ket{0,0,0}$. An electron tunnels from $L$ into $S_1$ at energy $\varepsilon_{s}^1$, followed by an electron tunneling into $G_1$ from $G$ at energy $\varepsilon_{g}+U_m$. At the next instant, the electron in $S_1$ tunnels into $S_2$, after which the electron in $G_1$ tunnels out into $G$ with energy $\varepsilon_g$. The cycle is completed and the system returns to the initial state when the electron in $S_2$ tunnels out into the reservoir $R$ with an energy $\varepsilon_s^2=\varepsilon_{s}^1+U_m$. It is clear that in this process an electron is transferred from $L$ to $R$ while absorbing a heat packet $U_m$ from $G$. Another cycle that again transfers electrons from $L$ to $R$, while absorbing heat packet from $G$ can be given by $\ket{0,0,0}\rightarrow \ket{1,0,0})\rightarrow\ket{1,1,0}\rightarrow\ket{0,1,1}\rightarrow\ket{0,0,1}\rightarrow\ket{0,0,0}$.  These transport processes contribute to thermoelectric power generation while absorbing heat energy from $G$ and are demonstrated as $(1)$  in Fig.~\ref{fig:Fig_6}. \color{black} Next, consider the cycle $\ket{0,0,0}\rightarrow \ket{1,0,0}\rightarrow\ket{1,1,0}\rightarrow\ket{0,1,0}\rightarrow\ket{0,0,0}$. This cycle consists of an electron tunneling into $S_1$ from $L$, with an energy $\varepsilon_{s}^1$, followed by an electron tunneling into $G_1$ with an energy $\varepsilon_g+U_m$. At the next step, the electron in $S_1$  exits into reservoir $L$ with an energy $\varepsilon_{s}^1+U_m$. The cycle is completed with the electron in $G_1$ tunnels out into $G$ with energy $\varepsilon_{g}$. It is evident that in this process, a packet of heat energy $U_m$ is transmitted from reservoir $G$ to $L$  without any net flow of electrons between $L$ and $R$. So, effectively the heat packet $U_m$ is wasted without any power conversion.  This electron flow component, depicted in Fig.~\ref{fig:Fig_6} as $(2)$, doesn't contribute to power generation, but transmits heat packets from $G$ and hence results in degradation of the efficiency. The third electron flow  component, depicted in Fig.~\ref{fig:Fig_6} as (3), results  from a reverse flux of electrons caused by the voltage bias (voltage drop across the load impedance) and flows in the direction opposite to the thermoelectric current. Hence, this component  results in the degradation of the net thermoelectric current and thus deteriorates the generated power as well as the efficiency. \color{black} Once the ground state of both $S_1$ and $G_1$ are  occupied, the electron existing in  $S_1$ can either tunnel into $S_2$ and subsequently into $R$ giving rise to  directional electronic flow (component 1) or tunnel out to $L$ without any net electron flow (component 2). Hence, the overall maximum power output of the proposed set-up hovers around $50\%$ of the optimal design. The generation efficiency at the overall maximum power is lower than $50\%$ of the optimal design due to component (3) of the total electron current. It should be noted that component (3) of the current only flows when the ground state of dot $G_1$ is already occupied with an electron and that of dot $S_1$ is empty.\\
\indent To further establish the points discussed above, in Fig.~\ref{fig:Fig_5}, I separate out the current flow into the system from the reservoir $L$ as:
\[
I_L=I_{L1}+I_{L2},
\]
where 
\begin{align}
&I_{L1}=q\gamma_c \times \left\{P^{\varsigma_1}_{0,0}f_L(\varepsilon_s^1)-P^{\varsigma_1}_{1,0}(1-f_L(\varepsilon_s^1)\right)\} \nonumber \\ 
&I_{L2}= q\gamma_c \{P^{\varsigma_1}_{0,1}f_L(\varepsilon_s^1+U_m)-  P^{\varsigma^1}_{1,1}\{1-f_L(\varepsilon_s^1+U_m)\}\} 
\label{eq:split}
\end{align}
 In Eq.~\eqref{eq:split}, $I_{L1}$ and $I_{L2}$ denote the total electronic current  from  reservoir $L$ into  the energy level $\varepsilon_s^1$ and  the Coulomb blockaded  level $\varepsilon_s^1+U_m$ respectively.  Fig.~\ref{fig:Fig_5}(a) and (b) demonstrate the electron current flow into the system from reservoir $L$ through the energy level $\varepsilon_s^1$ ($I_{L1}$) and  the Coulomb blockaded  level $\varepsilon_s^1+U_m$ ($I_{L2}$) respectively, while Fig.~\ref{fig:Fig_5}(c) demonstrates the total electronic current $I_L=I_{L1}+I_{L2}$ from the reservoir $L$ into the system. It should be noted that the electronic current demonstrated in Fig.~\ref{fig:Fig_5}, is opposite to the direction of conventional current flow (since electron charge is negative). We find that  the electron current flow $I_{L1}$ through $\varepsilon_{s}^1$, into the system from $L$, is positive or against the voltage bias, generating a net value of thermoelectric power. Interestingly, we also find that the current component $I_{L2}$ through the Coulomb blockaded energy level $\varepsilon_{s}^1+U_m$ is negative, as already shown in Fig.~\ref{fig:Fig_6}. That is, the net electron current, through the Coulomb blockaded level $\varepsilon_{s}^1+U_m$ flows into the reservoir $L$ from the system in the direction of voltage bias. This is already illustrated  in Fig.~\ref{fig:Fig_6}, where it is shown that the components $(2)$ and $(3)$ flow out of the system into $L$ via the Coulomb blocked level $\varepsilon_{s}^1+U_m$. As discussed above, these components $(2)$ and $(3)$ of electron  current flow may transmit heat packets without any power generation, or may reduce the net forward current flow against the voltage bias. Thus, they impact negatively on the generated power, as well as the efficiency.  The   deterioration in the heat engine performance, due to these current components $(2)$ and $(3)$ (discussed above), can be alleviated by adding an extra filter between $L$ and $S_1$, to restrict the current flow via the Coulomb blocked level $\varepsilon_s^1+U_m$, thereby nullifying the current components $(2)$ and $(3)$. However, doing so neutralizes the novelty of the proposed set-up in terms of fabrication simplicity. It should be noted that in Fig.~\ref{fig:Fig_5}(c),  negative values of total electronic current corresponds to a net electronic flow in the direction of the applied bias and thus the zone with negative electronic current flow  indicates the regime of \textit{zero} net thermoelectric power generation. \\
\section{Conclusion}\label{conclusion}
\indent To conclude, in this paper I have proposed a realistic design strategy for non-local heat engine based on Coulomb coupled systems. The performance of the proposed design  was then theoretically analyzed and compared with the optimal set-up \cite{coulomb_TE6} using the QME approach. It was demonstrated that the proposed set-up outputs a maximum power of around $50\%$ of  the optimal set-up. However, the crucial advantage of the proposed design strategy is that along with a reasonable output power, it also circumvents the demand for a sharp step-like change in reservoir-to-system coupling, which is required for proper operation of the optimal set-up proposed in literature\cite{coulomb_TE6}.  In the above discussions,  I have limited transport through the quantum dots in the weak coupling regime, so that the effects of co-tunneling can be neglected. The generated power in the proposed system can be increased by a few orders of magnitude by tuning electronic transport in the strong coupling regime, that is by increasing the system-to-reservoir, as well as, the interdot tunnel coupling. It will be interesting to study the effects of cotunneling on heat-harvesting performance of the proposed system as electronic transport is gradually tuned towards the strong coupling regime. \color{black}    In addition, I have also assumed ideal conditions by neglecting the lattice thermal conductance.  Understanding the  impact of  lattice heat flux on the generation efficiency and an investigation on the effect electron-phonon interaction  on the performance of the  proposed design also constitute interesting research directions. Although not shown here, the proposed system can also work as an efficient non-local heat engine when the reservoir $G$ acts as a heat sink (cold) with respect to the reservoirs $L$ and $R$ (hot). In such a case, the direction of thermoelectric current flow is reversed. The  different possible design strategies for non-local heat engines and their performance is left for future exploration. Nevertheless, the  set-up proposed in this paper  can be employed to fabricate high performance non-local heat engines using Coulomb coupled systems.\\
\indent \textbf{Acknowledgments:} The author would like to thank Sponsored Research and Industrial Consultancy (IIT Kharagpur) for their financial support via grant no. IIT/SRIC/EC/MWT/2019-20/162.\\
\indent \textbf{Data Availability:} The data that supports the findings of this study are available within the article
\appendix

\section{Another Coulomb coupled system based set-up for efficient non-local heat harvesting}\label{app_e}
\begin{figure}
	\centering
	\includegraphics[width=.45\textwidth]{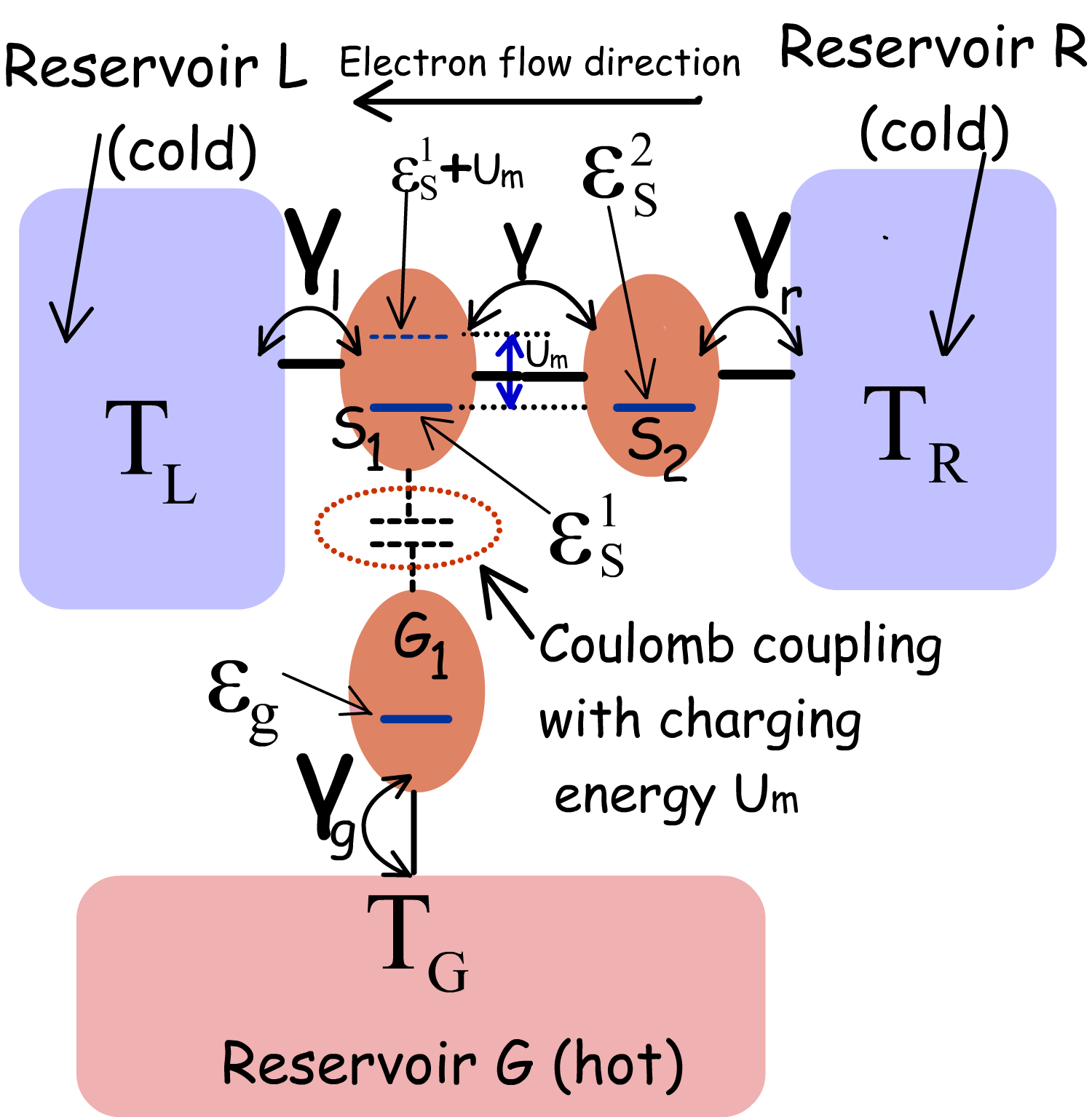}
	\caption{Schematic diagram for another realistic  design strategy to accomplish non-local heat harvesting using Coulomb coupled systems. The heat harvesting performance of this system was found to be similar to the proposed set-up in Fig.~\ref{fig:Fig_1} with slightly shifted regime of operation. }
	\label{fig:app_2}
\end{figure}

In this section, I show another possible set-up for non-local heat engine using Coulomb coupled systems that circumvents the need of a sharp transition in system-to-reservoir coupling around the ground states of the quantum dots. The system, demonstrated in Fig.~\ref{fig:app_2}, consists of three quantum-dots $S_1,~S_2$ and $G_1$, which are coupled to the reservoirs $L,~R$ and $G$ respectively. Identical to the set-up in Fig.~\ref{fig:Fig_1}, $S_1$ and $S_2$ are tunnel coupled to each other, while $G_1$ is Coulomb coupled to $S_1$ with electrostatic charging energy $U_m$. However, unlike the proposed set-up in Fig.~\ref{fig:Fig_1}, the ground-states of the two quantum dots $S_1$ and $S_2$ are identical, that is $\varepsilon_{s}^1=\varepsilon_{s}^2$. The asymmetry of the system ground state configuration, required for non local heat harvesting, creeps in when an electron tunnels from $G$ into $G_1$, such that an electron tunneling into (or out of) $S_1$ must now have an energy $\varepsilon_{s}^1+U_m$ when the ground state of $G_1$ is occupied. Although not detailed here, the system depicted in Fig.~\ref{fig:app_2} demonstrates similar heat-harvesting performance to the set-up shown in Fig.~\ref{fig:Fig_1} with slightly shifted regime of operation. In this case, for $T_G>T_{L(R)}$, the short-circuited electron flux, due to thermoelectric force, flows from reservoir $R$ to $L$, that is in the direction opposite to that of the set-up in Fig.~\ref{fig:Fig_1}. The set-up, demonstrated in Fig.~\ref{fig:app_2}, thus can also be employed for efficient non-local heat harvesting.
\section{Direction of heat flow in the Coulomb coupled systems.}\label{app_a}
In this section, I discuss the direction of electronic heat flow in a simple coulomb coupled system with constant energy-resolved system-to-reservoir coupling, for $T_G>T_{L(R)}$. Such a system is depicted in Fig.~\ref{fig:app_1}. Let us consider a sequential cycle that transfers a heat packet $U_m$ from the reservoir $G$ to the reservoir $L(R)$.  Consider the cycle,  demonstrated schematically in Fig.~\ref{fig:app_1}(b). With $n_x$ denoting the total number of electrons in the ground state of dot $x$ at any given instant of time, I denote each state of the complete system as $\ket{n_{S_1},n_{G_1}}$, where $n_{S_1},n_{G_1}\in (0,1)$. Consider the sequence of sequential electron transport ( Fig.~\ref{fig:app_1}b) given by $F_1\Rightarrow
 \ket{0,0}\rightarrow\ket{1,0)}\rightarrow\ket{1,1}\rightarrow\ket{0,1}\rightarrow\ket{0,0}$. In this cycle, the system starts with the initial vacuum state $\ket{0,0}$. An electron then tunnels from $L(R)$ into $S_1$ at energy $\varepsilon_{s}^1$, followed by an electron tunneling from $G$ into $G_1$ with an energy $\varepsilon_{g}+U_m$. Next, the  electron in $S_1$ tunnels out into $L(R)$ with an energy $\varepsilon_{s}^1+U_m$. The system returns to its ground state when the electron in $G_1$ tunnels out to $G$ with an energy $\varepsilon_{g}$. In this cycle, a heat packet is transferred from $G$ to $L(R)$.  The reverse cycle given by $R_1\Rightarrow \ket{0,0}\rightarrow\ket{0,1}\rightarrow \ket{1,1}\rightarrow \ket{1,0}\rightarrow \ket{0,0}$, on the other hand, transfers a heat energy $U_m$ from the reservoir $L(R)$ to $G$. In what follows, I show that that when $T_G>T_{L(R)}$, the probability of forward cycle $F_1$ (given in Fig.~\ref{fig:app_1}.b) is higher than than the probability of occurance of the reverse cycle, resulting in a net flow of heat from reservoir $G$ to $R$. If the energy resolved coupling of the reservoirs $L$ or $R$ to the dot $S_1$ is constant, then the electron in $S_1$ can tunnel in(out) from $L$ or $R$ with equal probability, thereby producing \textit{zero} net current.    Let us assume that the system, with schematic shown in Fig.~\ref{fig:app_1}(a), is initially in the vacuum state at time $t=0$ with no electrons in the levels $\varepsilon_g$ and $\varepsilon_s^1$. For simplicity, I also assume that there is no voltage drop between the reservoirs $L$ and $R$. Let us assume that the probability an electron enters the dot $S_1$ from the left contact between time $t$ and $t+dt$  is given by $P_{in}(t)dt$. We can write $P_{in}(t)dt$ as:
\begin{figure*}
	\centering
	\includegraphics[width=.95\textwidth]{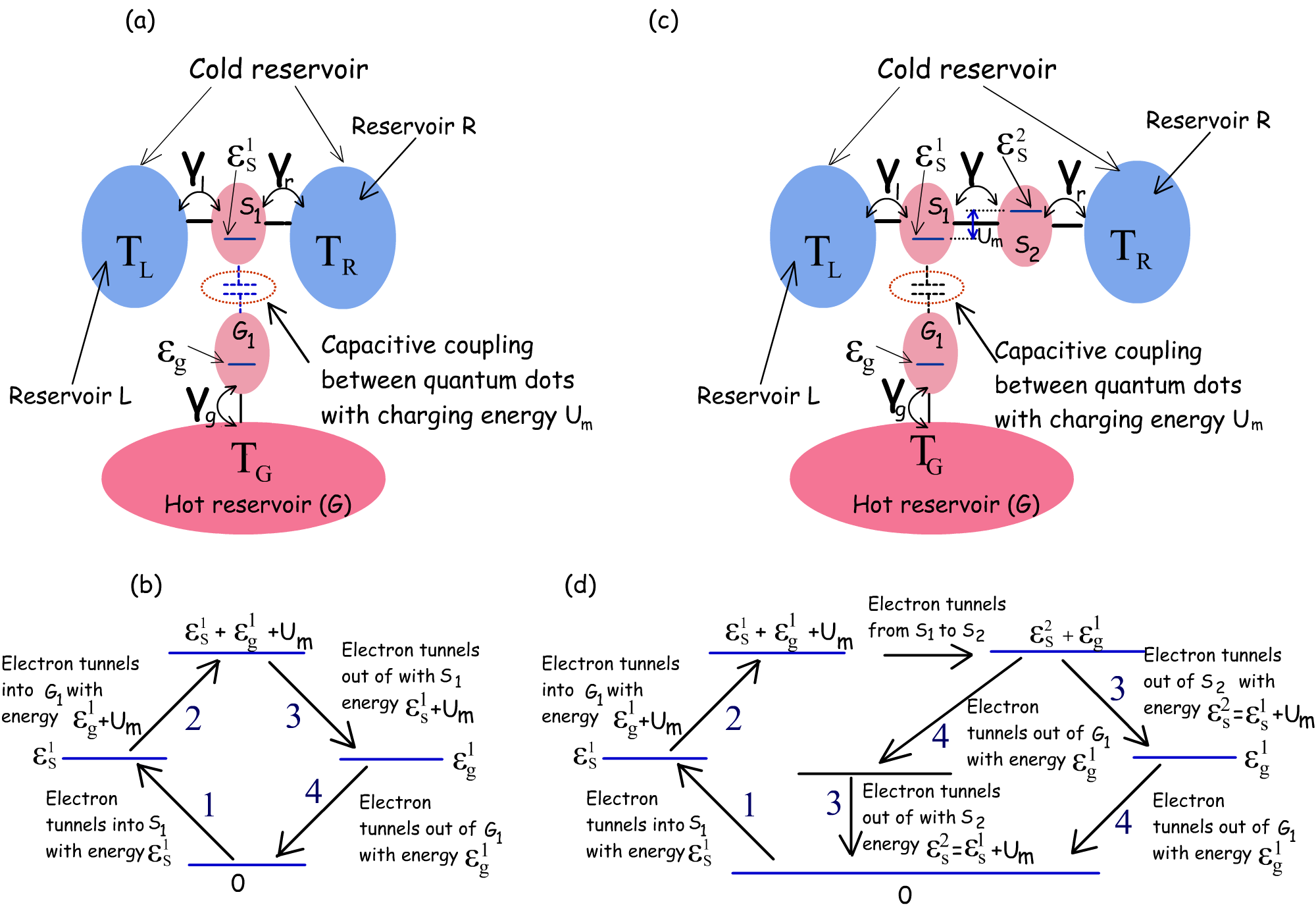}
	\caption{(a) Schematic of a Coulomb coupled  system consisting of a pair of quantum dots $S_1$ and $G_1$. The dot $S_1$ is electrically connected to the reservoirs $L$ and $R$, while $G_1$ is electrically connected to the reservoir $G$. $G_1$ and $S_1$ are Coulomb coupled with a mutual charging energy $U_m$. For constant energy resolved coupling $\gamma_l(\varepsilon)=\gamma_r(\varepsilon)=\gamma_g(\varepsilon)=\gamma_c$, the system produces no net thermoelectric current between $L$ and $R$. However, the stochastic thermal  fluctuation of $G$ can be converted into a directed thermoelectric current flow for asymmetric system-to-reservoir coupling given by $\gamma_l(\varepsilon)=\gamma_c \theta(\varepsilon_s^1+\delta \epsilon-\varepsilon),~\gamma_r(\varepsilon)=\gamma_c \theta(\varepsilon-\delta \epsilon-\varepsilon_s^1)$ and $\gamma_g(\varepsilon)=\gamma_c$, for $T_G\neq T_{L(R)}$ $(\delta \epsilon<{U_m}$) \cite{coulomb_TE1,coulomb_TE2,coulomb_TE3,coulomb_TE4,coulomb_TE5,coulomb_TE6}. (b) A sequential electron transport cycle that is responsible for heat packet flow from the reservoir $G$ to the reservoir $L(R)$. (c) Schematic of the realistic set-up proposed in this paper for efficient non-local heat harvesting. In this case, the system ground state configuration is asymmetric with respect to the reservoirs $L$ and $R$, with $\varepsilon_s^2=\varepsilon_{s}^1+\Delta\varepsilon$. In such a system, a non-local thermoelectric current can be driven between the reservoirs $L$ and $R$, by extracting heat from the reservoir $G$, even for the case of  constant energy resolved system-to-reservoir coupling given by  $\gamma_l(\varepsilon)=\gamma_r(\varepsilon)=\gamma_g(\varepsilon)=\gamma_c$. (d) Sequential transport processes leading to  transfer of electrons from $L$ to $R$ while absorbing heat from the reservoir $G$.}
	\label{fig:app_1}
\end{figure*}
\begin{equation}
P_{in}(t)dt=\gamma_c dt \times f_L(\varepsilon_s^1) \times \left\{1-\int_0^tP_{in}(\tau)d\tau\right\}
\end{equation}
Differentiating the above equation with respect to $t$, we get
\begin{equation}
\frac{dP_{in}(t)}{dt}=-\gamma_c  \times f_L(\varepsilon_s^1) \times P_{in}(t)
\end{equation}
Subject to the condition $\int_0^{\infty}P_{in}(t)dt=1$, the above equation has the solution
\begin{equation}
P_{in}(t)={\gamma_C f_L(\varepsilon_s^1)}e^{-\gamma_C f_L(\varepsilon_s^1)t}
\label{eq:P_in}
\end{equation}
Similarly, it can shown that an electron present in the energy level $\varepsilon_s^1$ at $t=0$ can exit to the left contact with a probability given by 
\begin{equation}
P_{out}(t)={\gamma_C \{1-f_L(\varepsilon_s^1)}\}e^{-\gamma_C \{1-f_L(\varepsilon_s^1)\}t}.
\label{eq:P_out}
\end{equation}
With the help of equations \eqref{eq:P_in} and \eqref{eq:P_out}, the probability that the forward cycle in Fig.~\ref{fig:app_1}(b) (marked with black arrows)  completes within a time-span $\Delta t$ can be written as:
\begin{widetext}
	\begin{eqnarray}
	\varrho_f=\int_0^{\Delta t}{\gamma_C \{f_L(\varepsilon_s^1)+f_R(\varepsilon_s^1)}\}e^{-\gamma_C \{f_L(\varepsilon_s^1)+f_R(\varepsilon_s^1)\}t_1}  
	\int_{t_1}^{\Delta t}{\gamma_C f_G(\varepsilon_g+U_m)}e^{-\gamma_C f_G(\varepsilon_g+U_m)t_2} \nonumber \\
	\times  \int_{t_2}^{\Delta t}{\gamma_C \{2-f_L(\varepsilon_s^1+U_m)-f_R(\varepsilon_s^1+U_m)}\}e^{-\gamma_C \{2-f_L(\varepsilon_s^1+U_m)-f_R(\varepsilon_s^1+U_m)\}t_3}  \nonumber \\
	\times  \int_{t_3}^{\Delta t}{\gamma_C\{1- f_G(\varepsilon_g)}\}e^{-\gamma_C \{1-f_G(\varepsilon_g)\}t_4}dt_4 dt_3 dt_3 dt_1
	\end{eqnarray}
	\begin{eqnarray}
	\Rightarrow \varrho_f=\Pi_f \int_0^{\Delta t}e^{-\gamma_C \{f_L(\varepsilon_s^1)+f_R(\varepsilon_s^1)\}t_1}  
	\int_{t_1}^{\Delta t}e^{-\gamma_C f_G(\varepsilon_g+U_m)t_2} 
	\int_{t_2}^{\Delta t}e^{-\gamma_C \{2-f_L(\varepsilon_s^1+U_m)-f_R(\varepsilon_s^1+U_m)\}t_3}  \nonumber \\
	\times  \int_{t_3}^{\Delta t}e^{-\gamma_C \{1-f_G(\varepsilon_g)\}t_4}dt_4 dt_3 dt_3 dt_1,
	\label{eq:varrho1}
	\end{eqnarray}
	\end{widetext}
	where $\Pi_f={\gamma_C^4 \{f_L(\varepsilon_s^1)+f_R(\varepsilon_s^1)}\} \times { f_G(\varepsilon_g+U_m)}\times { \{2-f_L(\varepsilon_s^1+U_m)-f_R(\varepsilon_s^1+U_m)}\} \times {\{1- f_G(\varepsilon_g)}\}$.
	To simplify the above equation, I work in the limit of very small $\Delta t$ such that the exponential quantities in the above equations is approximately 1. Hence, 
	\begin{align}
	\varrho_f & =\Pi_f \int_0^{\Delta t}
	\int_{t_1}^{\Delta t}
	\int_{t_2}^{\Delta t} 
	\int_{t_3}^{\Delta t}dt_4 dt_3 dt_3 dt_1 \nonumber \\
	& = \Pi_f \frac{\Delta t^4}{24}
	\end{align}
	Similarly, assuming that the system is in the vacuum state $\ket{0,0}$ at $t=0$, the probability that the reverse process is completed within a  very small time interval $\Delta t$ is
	\begin{align}
	\varrho_b & =\Pi_b \int_0^{\Delta t}
	\int_{t_1}^{\Delta t}
	\int_{t_2}^{\Delta t} 
	\int_{t_3}^{\Delta t}dt_4 dt_3 dt_3 dt_1 \nonumber \\
	& = \Pi_b \frac{\Delta t^4}{24},
	\end{align}
	where  $\Pi_b={\gamma_C^4 \{2-f_L(\varepsilon_s^1)-f_R(\varepsilon_s^1)}\} \times \{ 1-f_G(\varepsilon_g+U_m)\}
	\times { \{f_L(\varepsilon_s^1+U_m)+f_R(\varepsilon_s^1+U_m)}\} \times  f_G(\varepsilon_g)$.
	So, the ratio of the probability of occurance of the forward process to the reverse process for a very small instant of time $\Delta t$ is
	\begin{widetext}
	\begin{align}
	\varrho_f/\varrho_b & =\frac{\{f_L(\varepsilon_s^1)+f_R(\varepsilon_s^1)\}}{\{2-f_L(\varepsilon_s^1)-f_R(\varepsilon_s^1)\}}\times \frac{{ f_G(\varepsilon_g+U_m)}}{ \{ 1-f_G(\varepsilon_g+U_m)\}} \times \frac{ { \{2-f_L(\varepsilon_s^1+U_m)-f_R(\varepsilon_s^1+U_m)}\}}{{ \{f_L(\varepsilon_s^1+U_m)+f_R(\varepsilon_s^1+U_m)}\} }\times \frac{ {\{1- f_G(\varepsilon_g)}\}}{  f_G(\varepsilon_g)} \nonumber \\
	& = exp\left\{-\frac{\varepsilon_s^1-\mu}{kT_{L(R)}} \right\}exp\left\{-\frac{\varepsilon_g+U_m-\mu_g}{kT_G}\right\}  exp\left\{\frac{\varepsilon_s^1+U_m-\mu}{kT_{L(R)}} \right\} exp\left\{ \frac{\varepsilon_g-\mu_g}{kT_G}\right\} \nonumber \\
	& = exp\left\{\frac{U_m}{kT_{L(R)}} \right\}exp\left\{ -\frac{U_m}{kT_G}\right\} \nonumber \\
	& =exp\left\{\frac{U_m}{k}\left(\frac{1}{T_{L(R)}}-\frac{1}{T_G}\right)\right\}
	\label{eq:ratio_1}
	\end{align}
	\end{widetext}
	From Eq.~\eqref{eq:ratio_1}, it is clear that $\varrho_f=\varrho_b$ for $T_{L(R)}=T_G$. However, for $T_G>T_{L(R)}$, we have $\varrho_f>\varrho_b$, that is the probability of the forward process in Fig.~\ref{fig:app_1}(b) is greater than the probability of the reverse process. The same can be proved in the limit of finite and large $\Delta t$  by a analytical/numerical solution of Eq.~\eqref{eq:varrho1}. Hence, for $T_G>T_{L(R)}$, the direction of electronic heat flow via the Coulomb coupled system is from the reservoir $G$ to the reservoir(s) $L (R)$.
	\begin{widetext}
		\section{Direction of electron flow (for V=0) in the proposed set-up (demonstrated in Fig.~\ref{fig:Fig_1} or \ref{fig:app_1}c)}\label{app_c}
		Here, I show that for the arrangement demonstrated in Fig.~\ref{fig:Fig_1} or \ref{fig:app_1}(c), we get a directed motion of electrons from the reservoir $L$ to  $R$ under short-circuited condition for $T_G>T_{L(R)}$. Let us assume that the system is initially in the vacuum state $\ket{0,0,0}$. From Eqns.~\eqref{eq:P_in} and \eqref{eq:P_out}, the probability that the forward cycle, that transfers an electron from $L$ to $R$ (demonstrated in Fig.~\ref{fig:app_1}d) while absorbing a heat packet $U_m$ from the reservoir $G$, is completed in a time-span $\Delta t$ can be written as:
		\begin{align}
		\varrho_{L\rightarrow R}& =\int_0^{\Delta t}\gamma_C f_L(\varepsilon_s^1)e^{-\gamma_C f_L(\varepsilon_s^1)t_1}  
		\int_{t_1}^{\Delta t}{\gamma_C f_G(\varepsilon_g+U_m)}e^{-\gamma_C f_G(\varepsilon_g+U_m)t_2} \times \int_{t_2}^{\Delta t}\gamma e^{-\gamma t_3}   \nonumber \\
		&\times  \int_{t_3}^{\Delta t}{\gamma_C \{1-f_R(\varepsilon_s^2)}\}e^{-\gamma_C \{1-f_R(\varepsilon_s^2)\}t_4}  
		\times  \int_{t_3}^{\Delta t}{\gamma_C\{1- f_G(\varepsilon_g)}\}e^{-\gamma_C \{1-f_G(\varepsilon_g)\}t_5}dt_5dt_4 dt_3 dt_3 dt_1
		\end{align}
		\begin{eqnarray}
		\Rightarrow \varrho_{L\rightarrow R}  =\Pi_{L\rightarrow R} \int_0^{\Delta t}e^{-\gamma_C f_L(\varepsilon_s^1)t_1}  
		\int_{t_1}^{\Delta t}e^{-\gamma_C f_G(\varepsilon_g+U_m)t_2} \times \int_{t_2}^{\Delta t} e^{-\gamma t_3}   
		\times  \int_{t_3}^{\Delta t}
		e^{-\gamma_C \{1-f_R(\varepsilon_s^2)\}t_4}  \nonumber \\
		\times  \int_{t_3}^{\Delta t}e^{-\gamma_C \{1-f_G(\varepsilon_g)\}t_5}dt_5dt_4 dt_3 dt_3 dt_1
		\label{eq:varrhoLR}
		\end{eqnarray}
		where $\Pi_{L\rightarrow R}=\gamma\gamma_C^4 f_L(\varepsilon_s^1) \times { f_G(\varepsilon_g+U_m)}\times   \{1-f_R(\varepsilon_s^2)\} \times {\{1- f_G(\varepsilon_g)}\}$
		To simplify the above equation, I work in the limit of very small $\Delta t$ such that the exponential quantities in the above equations is approximately 1. Hence, 
		\begin{align}
		\varrho_{L\rightarrow R} & =\Pi_{L\rightarrow R} \int_0^{\Delta t}
		\int_{t_1}^{\Delta t}
		\int_{t_2}^{\Delta t} 
		\int_{t_3}^{\Delta t}\int_{t_3}^{\Delta t}dt_5dt_4 dt_3 dt_3 dt_1 \nonumber \\
		& = \Pi_{L\rightarrow R} \frac{\Delta t^5}{60}
		\label{eq:LR}
		\end{align}
		Similarly, assuming that the system is in ground state at $t=0$, the probability that the reverse cycle is completed within a  very small time interval $\Delta t$ is
		\begin{align}
		\varrho_{R\rightarrow L} & =\Pi_{R\rightarrow L} \int_0^{\Delta t}
		\int_{t_1}^{\Delta t}
		\int_{t_2}^{\Delta t} 
		\int_{t_3}^{\Delta t}\int_{t_3}^{\Delta t}dt_5dt_4 dt_3 dt_3 dt_1 \nonumber \\
		& = \Pi_{R\rightarrow L} \frac{\Delta t^5}{60},
		\end{align}
		where  $\Pi_{R\rightarrow  L}={\gamma \gamma_C^4 \{1-f_L(\varepsilon_s^1)}\} \times \{ 1-f_G(\varepsilon_g+U_m)\}
		\times f_R(\varepsilon_s^2) \times  f_G(\varepsilon_g)$.
		So, the ratio of the probability of the forward process to the reverse cycle to get completed within a very small interval of time $\Delta t$, is
		\begin{align}
		\varrho_{L \rightarrow R}/\varrho_{R \rightarrow L} & =\frac{\{f_L(\varepsilon_s^1)\}}{\{1-f_L(\varepsilon_s^1)\}}\times \frac{{ f_G(\varepsilon_g+U_m)}}{ \{ 1-f_G(\varepsilon_g+U_m)\}} \times \frac{ { \{1-f_R(\varepsilon_s^2)}\}}{{ \{f_R(\varepsilon_s^2)}\} }\times \frac{ {\{1- f_G(\varepsilon_g)}\}}{  f_G(\varepsilon_g)} \nonumber \\
		& = exp\left\{-\frac{\varepsilon_s^1-\mu}{kT_{L(R)}} \right\}exp\left\{-\frac{\varepsilon_g+U_m-\mu_g}{kT_G}\right\}  exp\left\{\frac{\varepsilon_s^2-\mu}{kT_{L(R)}} \right\} exp\left\{ \frac{\varepsilon_g-\mu_g}{kT_G}\right\} \nonumber \\
		& = exp\left\{\frac{U_m}{kT_{L(R)}} \right\}exp\left\{ -\frac{U_m}{kT_G}\right\} \nonumber \\
		& =exp\left\{\frac{U_m}{k}\left(\frac{1}{T_{L(R)}}-\frac{1}{T_G}\right)\right\}
		\label{eq:ratio_assy}
		\end{align}
		
	\end{widetext}
	For $T_G=T_{L(R)}$, $\varrho_{L \rightarrow R}=\varrho_{R \rightarrow L}$ and no electronic current flows between the reservoirs $L$ and $R$. However, for $T_G>T_{L(R)}$, we note that $\varrho_{L \rightarrow R}>\varrho_{R \rightarrow L}$, causing a net electronic current flow from the reservoir $L$ to resevoir $R$. The current reverses its direction for $T_G<T_{L(R)}$. \\
	\indent The variation in the thermoelectric power generation of the proposed set-up with $U_m$ (demonstrated in Fig.~\ref{fig:Fig_3}a) can be explained in terms of Eqns,~\ref{eq:LR}-\ref{eq:ratio_assy}. For low values of $U_m$,  the ratio $\varrho_{L \rightarrow R}/\varrho_{R \rightarrow L}$ is low which results in low values of the generated power. With an increase in $U_m$, the ratio $\varrho_{L \rightarrow R}/\varrho_{R \rightarrow L}$ increases and the forward cycle in Fig.~\ref{fig:app_1}(d), that generates power while harvesting heat from the reservoir $G$, dominates (over the reverse cycle) resulting in an overall increase in the net generated power. With further increase in $U_m$ beyond a certain point, despite of an increase in $\varrho_{L \rightarrow R}/\varrho_{R \rightarrow L}$, the absolute value of $\varrho_{L \rightarrow R}$ deteriorates due to decrease in the product $f_G(\varepsilon_g+U_m)\{1-f_G(\varepsilon_g)\}$, resulting in a decrease in the factor $\Pi_{L \rightarrow R}$. Hence, the generated power decreases with further increase in $U_m$.
	\section{Derivation of quantum master equations (QME) for the proposed system}\label{app_b}
	\begin{figure*}[!htb]
		\centering
		\includegraphics[width=.92\textwidth]{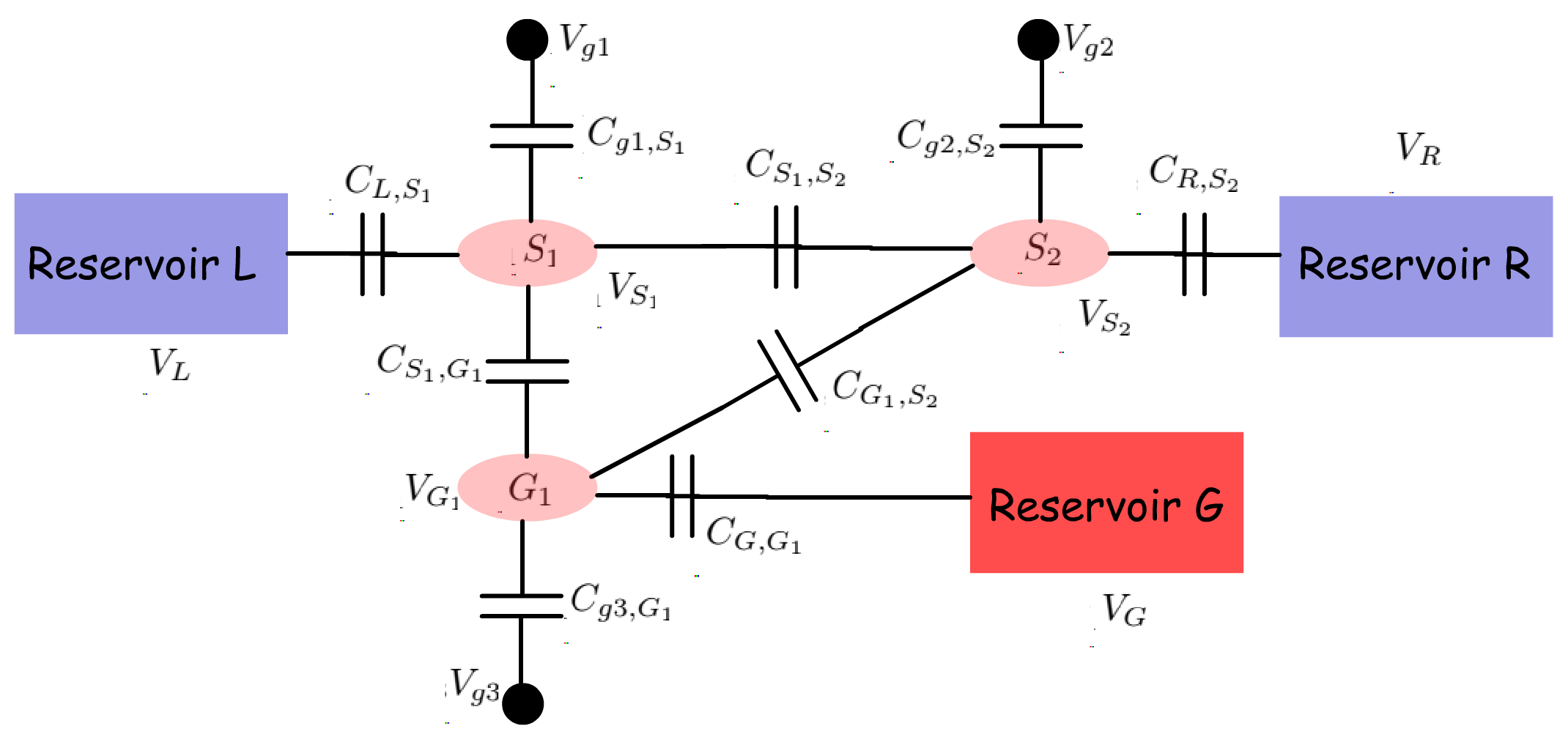}
		\caption{Schematic diagram demonstrating electrostatic interaction of the system with the adjacent electrodes and other dots. The voltages $V_{g1},~V_{g2},~V_{g3}$ are the voltages at the gate terminals of the dots $S_1$, $S_2$ and $G_1$ respectively. }
		\label{fig:app_5}
	\end{figure*}
Here, starting from the basic physics of Coulomb coupled systems, I derive the quantum master equations (QME) for the proposed non-local heat engine. Fig.~\ref{fig:app_5} demonstrates the equivalent model for electrostatic interaction of the system dots with the adjacent electrodes as well as the neighbouring dots.  The voltages $V_{g1},~V_{g2},~V_{g3}$ are the voltages at the gate terminals of the dots $S_1$, $S_2$ and $G_1$ respectively. The other symbols in Fig.~\ref{fig:app_5} are self-explanatory. The potentials of the dots $S_1,~S_2$ and $G_1$ can be written in terms of the dot charge and the neighbouring potentials  as \cite{cb}:
\begin{widetext}
	\begin{align}
&V_{S_1}=\frac{Q_{S_1}}{C_{S_1}^{\Sigma}}+\frac{1}{C_{S_1}^{\Sigma}}\left\{ C_{g1,S_1}V_{g1}+C_{L,S_1}V_{L}+C_{S_1,S_2}V_{S_2}+C_{S_1,G_1}V_{G_1}\right\} \nonumber \\
&V_{G_1}=\frac{Q_{G_1}}{C_{G_1}^{\Sigma}}+\frac{1}{C_{G_1}^{\Sigma}}\left\{ C_{g3,G_1}V_{g1}+C_{G,G_1}V_{G}+C_{G_1,S_2}V_{S_2}+C_{S_1,G_1}V_{S_1}\right\} \nonumber \\
&V_{S_2}=\frac{Q_{S_2}}{C_{S_2}^{\Sigma}}+\frac{1}{C_{S_2}^{\Sigma}}\left\{ C_{g2,S_2}V_{g2}+C_{R,S_2}V_{R}+C_{G_1,S_2}V_{G_1}+C_{S_1,S_2}V_{S_1}\right\} 
\label{eq:poential},
\end{align}
where $Q_x$ is the charge in dot $x$ and the terms $C_x^{\Sigma }$ is the total capacitance seen by the dot $x$ with its adjacent environment. 
\begin{align}
&C_{S_1}^{\Sigma }=C_{g1,S_1}+C_{S_1,G_1}+C_{L,S_1}+C_{S_1,S_2} \nonumber \\
&C_{S_2}^{\Sigma }=C_{g2,S_2}+C_{G_1,S_2}+C_{R,S_2}+C_{S_1,S_2} \nonumber \\
&C_{S_1}^{\Sigma }=C_{g13,G_1}+C_{S_1,G_1}+C_{G,G_1}+C_{G_1,S_2} 
\end{align}
 In general, each dot is strongly coupled with its corresponding gate. Hence, from a practical view-point, the effective capacitance $C_{L,S_1},~C_{R,S_2}$ and $C_{G,G_1}$ between the dots and the tunnel-coupled electrodes can be considered negligible with respect to the gate coupling capacitances $C_{g1,S_1},~C_{g2,S_2}$ and $C_{g3,G_1}$. In addition, the dots $S_1$ and $G_1$ are strongly coupled (intentionally) by suitable fabrication techniques \cite{cap_coup_1,cap_coup_2,cap_coup3,cap_coup_4,cap_coup_5}. The electrostatic coupling between $S_1-S_2$ and $G_1-S_2$ are again  negligible, such that for all practical purposes $C_{S_1,G_1}>>(C_{S_1,S_2},C_{G_1,S_2})$. For all practical purposes relating to the following derivations, I hence neglect the capacitances $C_{L,S_1},~C_{R,S_2},~C_{G,G_1},~C_{S_1,S_2},~C_{G_1,S_2}$. 
Under such considerations, the total electrostatic energy of the system can be written as \cite{cb}:
	\begin{align}
U_{tot}=\sum_{x\in (S_1,S_2,G_1)} \frac{Q_x^2}{2C_x^{\Sigma}} +\frac{Q_{S_1}}{C^{\Sigma}_{S_1}}\left\{C_{g1,S_1}V_{g1}+C_{S_1,G_1}V_{G_1}\right\}+\frac{Q_{S_2}}{C^{\Sigma}_{S_2}}C_{g2,S_2}V_{g2}+\frac{Q_{G_1}}{C^{\Sigma}_{G_1}}\left\{C_{g3,G_1}V_{g3}+C_{S_1,G_1}V_{S_1}\right\}, \nonumber \\
\label{eq:utot}
\end{align}
 where it is assumed that  capacitances $C_{L,S_1},~C_{R,S_2},~C_{G,G_1},~C_{S_1,S_2},~C_{G_1,S_2}\approx 0$, compared to the other capacitances in the system.
At $0K$ temperature, the system would reach equilibrium at the lowest possible value of $U_{tot}$, which we call $U_{eq}$. The  charge $Q_{S_1}^{eq}=-qn_{S_1}^{eq},~Q_{S_2}^{eq}=-qn_{S_2}^{eq}$ and $Q_{G_1}^{eq}=-qn_{G_1}^{eq}$ stored in the dot $S_1,~S_2$ and $G_1$ respectively in equilibrium (minimum energy) condition  at $0K$ can be found by solving the equations:
\begin{align}
&\frac{\partial U_{tot}}{\partial Q_{S_1}}=\frac{Q_{S_1}}{C^{\Sigma}_{S_1}}+\frac{1}{C_{S_1}^{\Sigma}}(C_{g1,S_1}V_{g1}+C_{S_1,G_1}V_{G_1})+\frac{(C_{S_1,G_1})^2 Q_{S_1}}{C_{G_1}^{\Sigma}(C_{S_1}^{\Sigma})^2}+\frac{C_{S_1,G_1}Q_{G_1}}{C_{S_1}^{\Sigma}C_{G_1}^{\Sigma}}=0 \nonumber \\
&\frac{\partial U_{tot}}{\partial Q_{S_2}}=\frac{Q_{S_2}}{C^{\Sigma}_{S_2}}+\frac{C_{g2,S_2}}{C_{S_2}^{\Sigma}}V_{g2} =0\nonumber \\
&\frac{\partial U_{tot}}{\partial Q_{G_1}}=\frac{Q_{G_1}}{C^{\Sigma}_{G_1}}+\frac{1}{C_{G_1}^{\Sigma}}(C_{g3,G_1}V_{g3}+C_{S_1,G_1}V_{G_3})+\frac{(C_{S_1,G_1})^2 Q_{G_1}}{C_{S_1}^{\Sigma}(C_{G_1}^{\Sigma})^2}+\frac{C_{S_1,G_1}Q_{S_1}}{C_{G_1}^{\Sigma}C_{S_1}^{\Sigma}}=0
\label{eq:partialderivative}
\end{align}
The above equations  have been derived by partial differentiation of Eq.~\ref{eq:utot}, in conjugation with replacing appropriate expressions by partially differentiating the set of Eqns.~\ref{eq:poential} along with some algebraic manipulation.
  At finite temperature, the number of electrons in the dots may vary stochastically  due to thermal fluctuations from the reservoir. The small increase in the total electrostatic energy of the system due to thermal fluctuations from the reservoirs can be written by Taylor expanding Eq.~\eqref{eq:utot} around the equilibrium dot charges ($-qn_{S_1}^{eq},~-qn_{S_2}^{eq}$ and $-qn_{G_1}^{eq}$), in conjugation with the condition $\frac{\partial U_{tot}}{\partial Q_{S_1}}\Big|_{Q_{S_1}=-qn_{S_1}^{eq}}=\frac{\partial U_{tot}}{\partial Q_{S_2}}\Big|_{Q_{S_2}=-qn_{S_2}^{eq}}=\frac{\partial U_{tot}}{\partial Q_{G_1}}\Big|_{Q_{G_1}=-qn_{G_1}^{eq}}=0$ as:
	\begin{eqnarray}
	U(n_{S_{1}},n_{G_{1}},n_{S_{2}})=U_{tot}-U_{eq}=\sum_{x \in (S_{1},G_{1},S_{2})}\frac{q^2}{C^{self}_{x}}\left(n_{x}^{tot}-n_x^{eq}\right)^2  +\sum_{(x_{1},x_{2})\in(S_{1},G_{1},S_{2})}^{x_1 \neq x_2} U_{x_1,x_2}\left(n_{x1}^{tot}-n_{x1}^{eq}\right)\left(n_{x2}^{tot}-n_{x2}^{eq}\right)  \nonumber \\
	\label{eq:caps}
	\end{eqnarray}
	\end{widetext}
	where $n_x^{tot}$ is the total number of electrons in the dot $x$, and    $C_x^{self}$ is self capacitance of dot $x$. $U_{x_1,x_2}$ is the electrostatic energy arising out of mutual  Coulomb coupling between two different quantum dots. These quantities can be derived from Eqns.~\eqref{eq:poential}, \eqref{eq:utot} and  \eqref{eq:partialderivative}, along with the assumption $C_{L,S_1}=C_{R,S_2}=C_{G,G_1}=C_{S_1,S_2}=C_{G_1,S_2}= 0$ as:
\begin{align}
&\frac{1}{C_{S_1}^{self}}=\frac{\partial^2 U_{tot}}{\partial Q_{S_1}^2}=\frac{1}{C_{S_1}^{\Sigma}}+2\frac{(C_{S_1,G_1})^2}{(C_{S_1}^{\Sigma})^2C_{G_1}^{\Sigma}} \nonumber \\
&\frac{1}{C_{S_2}^{self}}=\frac{\partial^2 U_{tot}}{\partial Q_{S_2}^2}=\frac{1}{C_{S_2}^{\Sigma}}\nonumber \\
&\frac{1}{C_{G_1}^{self}}=\frac{\partial^2 U_{tot}}{\partial Q_{G_1}^2}=\frac{1}{C_{G_1}^{\Sigma}}+2\frac{(C_{S_1,G_1})^2}{(C_{G_1}^{\Sigma})^2C_{S_1}^{\Sigma}} \nonumber \\
&\frac{U_{S_1,G_1}}{q^2}=\frac{\partial^2 U_{tot}}{\partial Q_{G_1}\partial Q_{S_1}}=\frac{\partial^2 U_{tot}}{\partial Q_{S_1}\partial Q_{G_1}}=2\frac{C_{S_1,G_1}}{C_{S_1}^{\Sigma}C_{G_1}^{\Sigma}} \nonumber
\end{align}
\begin{align}
&\frac{U_{S_1,S_2}}{q^2}=\frac{\partial^2 U_{tot}}{\partial Q_{S_1}\partial Q_{S_2}}=\frac{\partial^2 U_{tot}}{\partial Q_{S_2}\partial Q_{S_1}}=0\nonumber \\
&\frac{U_{S_2,G_1}}{q^2}=\frac{\partial^2 U_{tot}}{\partial Q_{G_1}\partial Q_{S_2}}=\frac{\partial^2 U_{tot}}{\partial Q_{S_2}\partial Q_{G_1}}=0
\label{eq:caps_expp}
\end{align}
	From Eq.~\eqref{eq:caps},  I proceed to derive the QME of the entire system. First, I consider the rate of inter-dot tunneling for the system proposed in this paper (schematically demonstrated in Fig.~\ref{fig:Fig_1} or \ref{fig:app_1}.c). In this arrangement, the additional quantum dot $S_2$ is tunnel coupled to the quantum dot $S_1$, while $G_1$ is Coulomb coupled to to $S_1$. I assume that the electrostatic energy arising out of  self-capacitance is much greater than the average thermal energy or the applied bias voltage, that is $E^{self}_x=\frac{q^2}{C^{self}_{x}}>> (kT,~qV)$, such that transport through the Coulomb blocked energy level, due to self-capacitance, is negligible. Under these assumptions, the analysis may be restricted to eight multi-electron  states, that can be denoted by the electron numbers in each quantum dot as $\ket{n_{S_1},n_{G_1},n_{S_2}}=\ket{n_{S_1}}\tens{} \ket{n_{G_1}} \tens{} \ket{n_{S_2}}$, where $(n_{S_1},n_{G_1},n_{S_2})\in (0,1)$. For simplifying the representation of these multi-electron states, I rename the states as $\ket{0,0,0}\rightarrow \ket{0}$, $\ket{0,0,1}\rightarrow \ket{1}$, $\ket{0,1,0}\rightarrow \ket{2}$, $\ket{0,1,1}\rightarrow \ket{3}$, $\ket{1,0,0}\rightarrow \ket{4}$, $\ket{1,0,1}\rightarrow \ket{5}$, $\ket{1,1,0}\rightarrow \ket{6}$, and 
	$\ket{1,1,1}\rightarrow \ket{7}$ \\
	The Hamiltonian of the entire system consisting of these three quantum dots, hence, can be written as:
	\begin{eqnarray}
	H=\sum_{\beta}\epsilon_{\beta}\ket{\beta}\bra{\beta}+t\{\ket{3}\bra{6}+\ket{1}\bra{4}\}\nonumber \\ +U_m \{\ket{6}\bra{6}+\ket{7}\bra{7}\}+h.c.,
	\end{eqnarray}
	where $U_m=U^m_{S_1,G_1}$ is the electrostatic coupling energy between $S_1$ and $G_1$ in Fig.~\ref{fig:Fig_1} (or \ref{fig:app_1}c), $t$ is the interdot tunnel coupling element between $S_1$ and $S_2$ and $\epsilon_{\beta}$ is the total energy of the state $\ket{\beta}$ with respect to the vacuum state $\ket{0}$.  Under the assumption that the interdot coupling element $t$ or the reservoir to dot coupling are small, the time evolution of the system density matrix can be calculated by taking the partial trace over the total density matrix of the combined set-up consisting of the reservoirs and the dots \cite{master_eq_1,master_eq_2,master_eq_3,master_eq_4,master_eq_5,master_eq_6}. In particular,  the diagonal and the non-diagonal terms of the density matrix $\rho$ of the system consisting only of the arrays of quantum dots may be written as a set of modified Liouville euqation \cite{master_eq_1,master_eq_2,master_eq_3,master_eq_4,master_eq_5,master_eq_6}:
	\begin{eqnarray}
	\frac{\partial \rho_{\eta \eta}}{\partial t}=-i[H,\rho]_{\eta \eta}-\sum_{\nu} \Gamma_{\eta \nu}\rho_{\eta \eta}+\sum_{\delta}\Gamma_{\delta \eta }\rho_{\delta \delta} \nonumber \\
	\frac{\partial \rho_{\eta \beta}}{\partial t}=-i[H,\rho]_{\eta \beta}-\frac{1}{2}\sum_{\nu} \Big(\Gamma_{\eta \nu}+\Gamma_{\beta \nu }\Big) \rho_{\eta \beta}, \nonumber \\
	\label{eq:time_derivate}
	\end{eqnarray}
	where $[x,y]$ denotes the commutator of the operators $x$ and $y$ and $\rho_{\eta \beta}=\bra{\eta}\rho \ket{\beta}$. The elements $\rho_{\eta \eta}$ and $\rho_{\eta \beta}$ in the above equation denote any diagonal and non-diagonal element of the system density matrix respectively. The off-diagonal elements $\rho_{\eta \beta}$ result from coherent inter-dot tunnelling and tunnelling of electrons between the dots and the reservoirs. The off-diagonal terms, hence, are only non-zero when electron tunneling can result in the transition between the states $\eta$ and $\beta$. The parameters $\Gamma_{xy}$ account for the transition between system   states  due to electronic tunneling between the system and the reservoirs and are only finite when the system state transition from  $\ket{x}$ to $\ket{y}$ (or vice-versa)  is possible due to electron transfer between the system and the reservoirs. Assuming  a statistical quasi-equilibrium distribution of electrons inside the reservoirs,  $\Gamma_{xy}$  can be expressed as:   
	\begin{eqnarray}
	\Gamma_{xy }=\gamma_{c}f_{\lambda}(\epsilon_{y}-\epsilon_{x}), 
	\label{eq:c4}
	\end{eqnarray}
	where ${\gamma_{c}=\gamma_l=\gamma_r=\gamma_g}$ denotes the system to reservoir coupling corresponding to the state transition,  $f_{\lambda}(\epsilon)$ denotes the probability of occupancy of an electron in the corresponding reservoir $\lambda$ (driving the state transition) at energy $\epsilon$ and $\epsilon_{x (y)}$ is the total electronic energy of the system in the state $\ket{x(y)}$ compared to the vacuum state. In the context of this discussion, I assume equilibrium Fermi-Dirac carrier statistics in the reservoir. \\
 For this system, the tunneling of electrons between the quantum dots corresponds to the change of the system states from $\ket{4}$ to $\ket{1}$ (or vice-versa) and from $\ket{3}$ to $\ket{6}$ (or vice-versa). In steady state, the time-derivative of each element of the density matrix $[\rho]$ vanishes. Hence, using the second equation of \eqref{eq:time_derivate}, we get, 
	\begin{equation}
	\rho_{4,1}=\rho^{*}_{1,4}=\frac{\rho_{4,4}-\rho_{1,1}}{\epsilon_{4}-\epsilon_{1}-i\frac{\Upsilon_{4,1}}{2}}
	\label{eq:c5}
	\end{equation}
	\begin{equation}
	\rho_{6,3}=\rho^{*}_{3,6}=\frac{\rho_{6,6}-\rho_{3,3}}{\epsilon_{6}-\epsilon_{3}-i\frac{\Upsilon_{6,3}}{2}},
	\label{eq:c6}
	\end{equation}
	where $\Upsilon_{x,y}$ is the sum of all process rates arising from the system to reservoir coupling that leads to the decay of the states $\ket{x}$ and $\ket{y}$. In Eqns.~\eqref{eq:c5} and \eqref{eq:c6}, $\Upsilon_{4,1}$ and $\Upsilon_{6,3}$ can be written as:
	\begin{gather}
	\Upsilon_{4,1}=\Gamma_{\ket{4},\ket{0}}+\Gamma_{\ket{4},\ket{6}}+\Gamma_{\ket{4},\ket{5}}+\Gamma_{\ket{1},\ket{0}}+\Gamma_{\ket{1},\ket{6}}+\Gamma_{\ket{1},\ket{3}} \nonumber \\
	\Upsilon_{6,3}=\Gamma_{\ket{6},\ket{4}}+\Gamma_{\ket{6},\ket{2}}+\Gamma_{\ket{6},\ket{7}}+\Gamma_{\ket{3},\ket{1}}+\Gamma_{\ket{3},\ket{2}}+\Gamma_{\ket{3},\ket{7}} 
	\end{gather}
	From the first equation of \eqref{eq:time_derivate}, the time derivative of the diagonal elements $\rho_{6,6}$ and $\rho_{3,3}$ of the density matrix can be written as:
	\begin{align}
	\dot{\rho}_{6,6}=&it(\rho_{6,3}-\rho_{3,6})-\left(\Gamma_{\ket{6},\ket{4}}-\Gamma_{\ket{6},\ket{2}}+\Gamma_{\ket{6},\ket{7}}\right)\rho_{6,6}\nonumber \\&+\Gamma_{\ket{4},\ket{6}}\rho_{4,4}+\Gamma_{\ket{2},\ket{6}}\rho_{2,2}+\Gamma_{\ket{7},\ket{6}}\rho_{7,7} \nonumber \\
	\dot{\rho}_{4,4}=&it(\rho_{4,1}-\rho_{1,4})-\left(\Gamma_{\ket{4},\ket{0}}-\Gamma_{\ket{4},\ket{6}}+\Gamma_{\ket{4},\ket{5}}\right)\rho_{4,4}\nonumber \\&+\Gamma_{\ket{0},\ket{4}}\rho_{0,0}+\Gamma_{\ket{6},\ket{4}}\rho_{6,6}+\Gamma_{\ket{5},\ket{4}}\rho_{5,5} \nonumber \\
	\label{eq:b8}
	\end{align}
	Substituting the values of  $\rho_{6,3},~\rho_{3,6},~\rho_{4,1}$ and $\rho_{1,4}$ in Eq.~\eqref{eq:b8}   from Eq.~\eqref{eq:c5} and \eqref{eq:c6}, the probability of decay of the states $\ket{4}$ and $\ket{6}$ can be written as:
	\begin{eqnarray}
	\dot{p_6}=\dot{\rho}_{6,6}=\sum_{\alpha=}\left(  -\Gamma_{\ket{6},\ket{\alpha}}p_6+\Gamma_{\ket{\alpha},\ket{6}}p_{\alpha} \right) \nonumber \\
	-\Lambda_{\ket{6},\ket{3}}p_6+\Lambda_{\ket{3},\ket{6}}p_3
	\end{eqnarray}\\
	\begin{eqnarray}
	\dot{p_4}=\dot{\rho}_{4,4}=\sum_{\alpha=}\left(  -\Gamma_{\ket{4},\ket{\alpha}}p_3+\Gamma_{\ket{\alpha},\ket{4}}p_{\alpha} \right) \nonumber \\
	-\Lambda_{\ket{4},\ket{1}}p_4+\Lambda_{\ket{1},\ket{4}}p_1,
	\end{eqnarray}
	where $p_{\eta}=\rho_{\eta,\eta}$ and 
	\begin{gather}
	\Lambda_{\ket{6},\ket{3}}=\Lambda_{\ket{3},\ket{6}}=t^2\frac{\Upsilon_{6,3}}{(\epsilon _6-\epsilon _3)^2+\frac{\Upsilon _{6,3}^2}{4}}\nonumber \\
	\Lambda_{\ket{4},\ket{1}}=\Lambda_{\ket{1},\ket{4}}=t^2\frac{\Upsilon_{4,1}}{(\epsilon _4-\epsilon _1)^2+\frac{\Upsilon _{4,1}^2}{4}} \nonumber \\
	\label{eq:tun_rate}
	\end{gather}
	In the above equation, $\Lambda_{\ket{4},\ket{1}}$ and $\Lambda_{\ket{6},\ket{3}}$ are the interdot tunneling rates when the ground state of  $G_1$ is unoccupied and occupied respectively. By a clever choice of the energy states, such that, $\epsilon_6=\varepsilon_g+\varepsilon_s^1+U_m=\varepsilon_g+\varepsilon_s^2=\epsilon_3$, that is by choosing $\varepsilon_s^2=\varepsilon_s^1+U_m$, we can arrive at a condition where $\Lambda_{\ket{6},\ket{3}}>>\Lambda_{\ket{4},\ket{1}}$ (under the condition $U_m>>|\Upsilon _{4,1}|$). Such a condition implies that the tunneling probability between the dots is negligible in the absence of an electron in $G_1$, which initiates a unidirection flow of electrons when $T_G>T_{L(R)}$.\\
\indent For the calculation of current, it is sufficient to know the probability of occupancy of the dot $S_1$ or $S_2$. Since, the electronic transport in $S_1$ and $G_1$ are coupled to each other via Coulomb interaction, I will treat $S_1$ and $G_1$ as a separate sub-system ($\varsigma_1$) of the entire system, $S_2$ being another sub-system ($\varsigma_2$) of the entire system consisting of the three dots. To simplify my calculations, I assume that $\Lambda_{\ket{4},\ket{1}}<<\Lambda_{\ket{6},\ket{3}}$, such that for all practical purposes relating to electron transport  $\Lambda_{\ket{4},\ket{1}}\approx 0$. In what follows, I simply denote $\Lambda_{\ket{6},\ket{3}}$ as $\gamma$. $\gamma$ thus denotes the interdot tunnel coupling. I write the probability of occupancy of the subsystem $\varsigma_1$ as $P_{i,j}^{\varsigma_1}$, where $i$ and $j$ denote the number of electrons in the ground state of the dot $S_1$ and $G_1$ respectively. $P_z^{\varsigma_2}$, on the other hand, would be used to denote the probability of occupancy of the dot $S_2$. Note that breaking down the entire system into two sub-system in this fashion is permissible only in the limit of weak tunnel and Coulomb coupling between the two sub-systems so that the state of one sub-system doesn't affect the state of the complementary sub-system. In such a limit, we can write $\rho_{0,0}=P^{\varsigma_1}_{0,0}P^{\varsigma_2}_{0},~\rho_{1,1}=P^{\varsigma_1}_{0,0}P^{\varsigma_2}_{1},~\rho_{2,2}=P^{\varsigma_1}_{0,1}P^{\varsigma_2}_{0},~\rho_{3,3}=P^{\varsigma_1}_{0,1}P^{\varsigma_2}_{1},~\rho_{4,4}=P^{\varsigma_1}_{1,0}P^{\varsigma_2}_{0},~\rho_{5,5}=P^{\varsigma_1}_{1,0}P^{\varsigma_2}_{1},~\rho_{6,6}=P^{\varsigma_1}_{1,1}P^{\varsigma_2}_{1},~\rho_{7,7}=P^{\varsigma_1}_{1,1}P^{\varsigma_2}_{1}$ The rate equations for the sub-system $\varsigma_1$ can be written in terms of two or more diagonal elements of the density matrix, in \eqref{eq:time_derivate}, as \cite{sispad}:
\begin{widetext}
	\begin{align}
	\frac{d}{dt}(P_{0,0}^{\varsigma_1})=\frac{d}{dt}\left( \rho_{0,0}+\rho_{1,1}\right)=&\gamma_c \times \left\{-P_{0,0}^{\varsigma_1}\{f_L(\varepsilon_s^1)+f_G(\varepsilon_g)\}+P_{0,1}^{\varsigma_1}\{1-f_G(\varepsilon_g)\}+P_{1,0}^{\varsigma_1}\{1-f_L(\varepsilon_s^1)\}\right\}\nonumber \\
	\frac{d}{dt}(P_{1,0}^{\varsigma_1})=\frac{d}{dt}\left(\rho_{5,5}+\rho_{4,4}\right)=&\gamma_c \times \left\{-P_{1,0}^{\varsigma_1}\left\{1-f_L(\varepsilon_{s}^1)+f_G(\varepsilon_g+U_m)\right\}+P_{1,1}^{\varsigma_1}\left\{1-f_G(\varepsilon_g+U_m)\right\}+P_{0,0}^{\varsigma_1}f_G(\varepsilon_g)\right\} \nonumber \\
	\frac{d}{dt}(P_{0,1}^{\varsigma_1})=\frac{d}{dt}\left(\rho_{2,2}+\rho_{3,3}\right)=&\gamma_c \times \left\{-P_{0,1}^{\varsigma_1}\left\{1-f_g(\varepsilon_{g}^1)+f_L(\varepsilon_s^1+U_m)+\frac{\gamma}{\gamma_c}P^{\varsigma_2}_1\right\}\right\} \nonumber \\& +\gamma_c \left\{P_{0,0}^{\varsigma_1}f_G(\varepsilon_g)+P_{1,1}^{\varsigma_1}\left\{1-f_L(\varepsilon_s^1+U_m)+\frac{\gamma}{\gamma_c}P^{\varsigma_2}_{0}\right\}\right\} \nonumber \\
	\frac{d}{dt}(P_{1,1}^{\varsigma_1})=\frac{d}{dt}\left(\rho_{7,7}+\rho_{6,6}\right)=&\gamma_c \times \left\{-P_{1,1}^{\varsigma_1}\left\{[1-f_g(\varepsilon_{g}^1+U_m)]+[1-f_L(\varepsilon_s^1+U_m)]+\frac{\gamma}{\gamma_C}P^{\varsigma_2}_0\right\}\right\} \nonumber \\ &+\gamma_c \left\{P_{1,0}^{\varsigma_1}f_G(\varepsilon_g+U_m) +P_{0,1}^{\varsigma_1}\left\{f_L(\varepsilon_s^1+U_m)+\frac{\gamma}{\gamma_c}P^{\varsigma_2}_{1}\right\}\right\} \nonumber \\
	\label{eq:first_sys1}
	\end{align} 
	where  $\Lambda_{\ket{4},\ket{1}}=\Lambda_{\ket{1},\ket{4}}=0$ and $\gamma=\Lambda_{\ket{6},\ket{3}}=\Lambda_{\ket{3},\ket{6}}$.  I  assume quasi Fermi-Dirac carrier statistics at the reservoirs such that $f_{\lambda}(\epsilon)=\left\{1+exp\left(\frac{\epsilon-\mu_{\lambda}}{kT_{\lambda}}\right)\right\}^{-1}$, corresponding to the reservoir $\lambda$, and $\lambda \in (L,R,G)$.
	Similarly, the rate equations of the sub-system $\varsigma_2$ can be written as:
	\begin{align}
	&\frac{d}{dt}(P_{0}^{\varsigma_2})=\frac{d}{dt}\left( \rho_{6,6}+\rho_{4,4}+\rho_{2,2}+\rho_{0,0}\right)=\gamma_c \times \left\{-P_{0}^{\varsigma_2}\{f_R(\varepsilon_s^2)+\frac{\gamma}{\gamma_c}P_{1,1}^{\varsigma_1}\}+P_1^{\varsigma_2}\{1-f_R(\varepsilon_{s}^2)+\frac{\gamma}{\gamma_c}P^{\varsigma_1}_{0,1}\}\right\}\nonumber \\
	&\frac{d}{dt}(P_{1}^{\varsigma_2})=\frac{d}{dt}\left( \rho_{7,7}+\rho_{5,5}+\rho_{3,3}+\rho_{1,1}\right)=\gamma_c \times \left\{-P_1^{\varsigma_2}\{1-f_R(\varepsilon_{s}^2)+\frac{\gamma}{\gamma_c}P^{\varsigma_1}_{0,1}\}+P_{0}^{\varsigma_2}\{f_R(\varepsilon_s^2)+\frac{\gamma}{\gamma_c}P_{1,1}^{\varsigma_1}\}\right\}\nonumber \\
	\label{eq:second_sys1}
	\end{align}
	In steady state, the L.H.S of Eqns.~\eqref{eq:first_sys1} and \eqref{eq:second_sys1} are zero. 	Both the sets of Eqns.~\eqref{eq:first_sys1} and \eqref{eq:second_sys1} form  dependent sets of equations. The dependency of  the sets of Eqns.~\eqref{eq:first_sys1} and \eqref{eq:second_sys1} is broken by introducing the probability conservation rules, that is, $\sum_{x,y}P_{x,y}^{\varsigma_1}=1 $ for ~\eqref{eq:first_sys1} and  $\sum_{z}P_{z}^{\varsigma_2}=1 $ for \eqref{eq:second_sys1}. The set of Eqns.~\eqref{eq:first_sys1} and \eqref{eq:second_sys1} form a coupled system of equations which, for the case of my study, were solved using iterative Newton-Raphson method. On solution of the state probabilities given by Eqns.~\eqref{eq:first_sys1} and \eqref{eq:second_sys1}, the charge current $I_{L(R)}$ through the system  and the electronic heat current ($I_{Qe}$) extracted from the reservoir $G$ can be calculated using the equations:
	\begin{equation}
	I_L= q\gamma_c \times \left\{P^{\varsigma_1}_{0,0}f_L(\varepsilon_s^1)+P^{\varsigma_1}_{0,1}f_L(\varepsilon_s^1+U_m)-P^{\varsigma_1}_{1,0}\{1-f_L(\varepsilon_s^1)\}-P^{\varsigma_s^1}_{1,1}\{1-f_L(\varepsilon_s^1+U_m)\}\right\} 
	\end{equation}
	\begin{equation}
	I_R= -q\gamma_c \times \left\{P^{\varsigma_2}_{0}f_R(\varepsilon_s^1)-P^{\varsigma_2}_{1}\{1-f_R(\varepsilon_s^1)\}\right\} 
	\end{equation}
	\begin{align}
	I_{Qe}=
	\gamma_c \times \left\{(\varepsilon_g+U_m-\mu_g)\left\{P^{\varsigma_1}_{1,0}f_G(\varepsilon_g+U_m)-P^{\varsigma_1}_{1,1}\{1-f_G(\varepsilon_g+U_m)\}\right\} \right\}\nonumber \\
	+ \gamma_c \times\left\{(\varepsilon_g-\mu_g)\times \left\{P^{\varsigma_1}_{0,0}f_G(\varepsilon_g)  -P^{\varsigma_n}_{0,1}\{1-f_G(\varepsilon_g)\}\right\}\right\}  
	\label{eq:heat2}
	\end{align}
	Since, no net current flows from the reservoir $G$, we have 
	\begin{equation}
	I_G=
	q\gamma_c \times \left\{P^{\varsigma_1}_{1,0}f_G(\varepsilon_g+U_m)-P^{\varsigma_1}_{1,1}\{1-f_G(\varepsilon_g+U_m)\}+P^{\varsigma_1}_{0,0}f_G(\varepsilon_g)  -P^{\varsigma_n}_{0,1}\{1-f_G(\varepsilon_g)\}\right\}=0
	\label{eq:gate_curr}
	\end{equation}
	Substituting Eq.~\eqref{eq:gate_curr} in Eq.~\eqref{eq:heat2}, the equation for $I_{Qe}$ becomes modified as:
	\begin{equation}
	I_{Qe}=\gamma_c \times U_m\left\{P^{\varsigma_1}_{1,0}f_G(\varepsilon_g+U_m)-P^{\varsigma_1}_{1,1}\{1-f_G(\varepsilon_g+U_m)\}\right\} 
	\end{equation}
	\end{widetext}
\section{Effect of the ratio $r=\gamma/\gamma_c$ on heat harvesting properties of the proposed set-up}\label{app_f}
\begin{figure}[!htb]
	\centering
	\includegraphics[width=.5\textwidth]{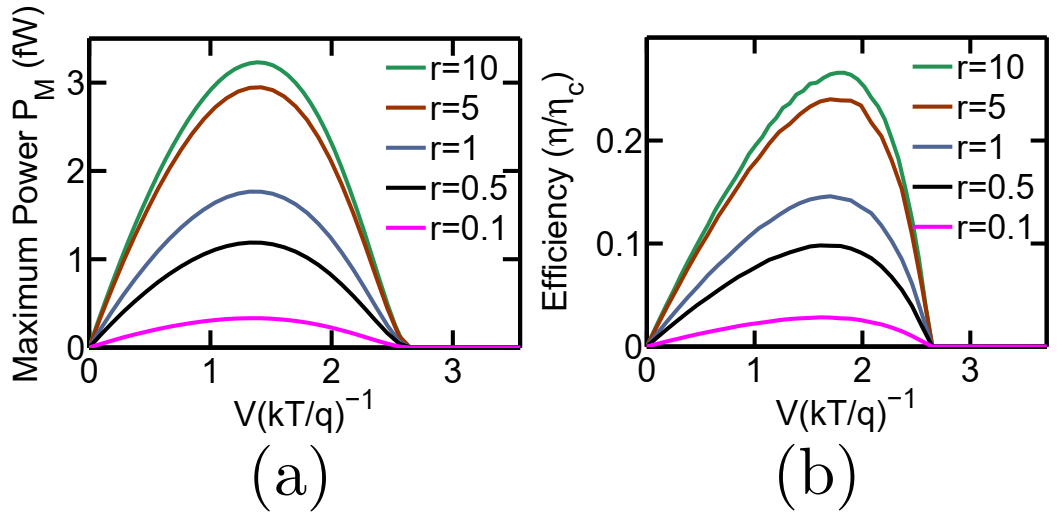}
	\caption{Variation in heat harvesting performance of the proposed system as  the ratio $r=\gamma/\gamma_c$ is tuned, keeping the system-to-reservoir coupling  constant at $\gamma_c=10^{-5}\frac{q}{h}$.  Plot of (a) maximum generated power ($P_M$) and (b) Efficiency at the maximum generated power, for various values of $r$, with variation in the applied voltage for $U_m=3.9meV(\approx 6kT/q)$ and $T_G=10K,~T_{L(R)}=5K$. $T=\frac{T_G+T_{L(R)}}{2}=7.5K$ is the average temperature of the system.}
	\label{fig:app_3}
\end{figure}
\begin{figure}[!htb]
	\centering
	\includegraphics[width=.5\textwidth]{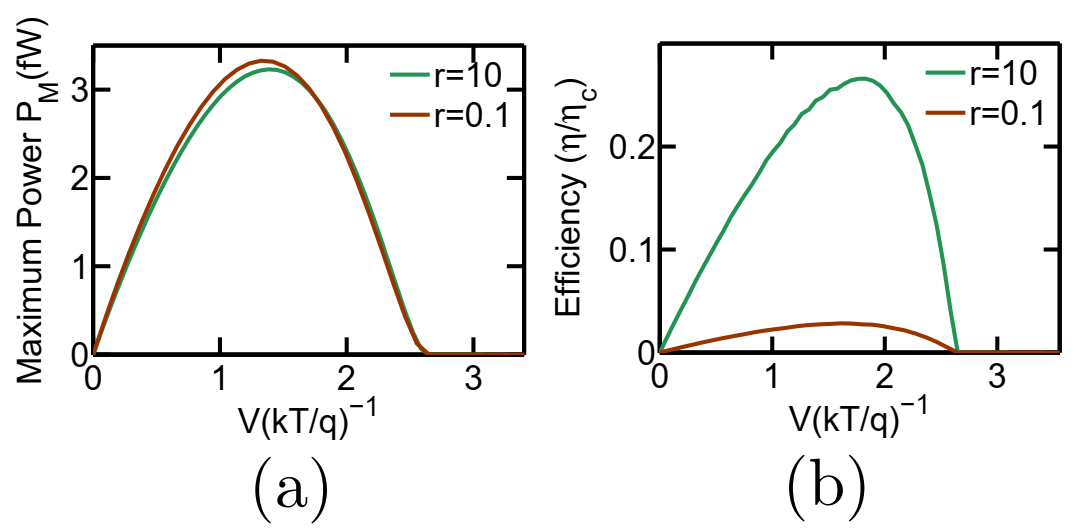}
	\caption{Variation in heat harvesting properties of the proposed system as the values of $\gamma_c$ and $\gamma$ are interchanged. The two cases demonstrated here are (i) $\gamma_c=10^{-5}\frac{q}{h}$ and $\gamma=10^{-4}\frac{q}{h}$, resulting in $r=10$ (shown in green) and (ii)  $\gamma_c=10^{-4}\frac{q}{h}$ and $\gamma=10^{-5}\frac{q}{h}$, resulting in $r=0.1$ (shown in brown). Plot of (a) maximum generated power ($P_M$) and (b) Efficiency at the maximum generated power,  with variation in the applied voltage for $U_m=3.9meV(\approx 6kT/q)$ and $T_G=10K,~T_L(R)=5K$. $T=\frac{T_G+T_{L(R)}}{2}=7.5K$ is the average temperature of the system.}
	\label{fig:app_4}
\end{figure}
In this section, I demonstrate the variation in heat harvesting properties of the proposed set-up with a variation in $r=\gamma/\gamma_c$. Throughout the main discussion in this paper, the values of the system-to-reservoir and interdot coupling were chosen to be $\gamma_c=10^{-5}\frac{q}{h}$ and $\gamma=10^{-4}\frac{q}{h}$.  Since, the conductance between the reservoirs $L$ and $R$ depends on both the interdot coupling and system-to-reservoir coupling , reducing either of the two deteriorates the conductance and hence, degrades the generated power. In the Fig.~\ref{fig:app_3} below, I plot the variation in maximum generated power, as well as the efficiency at the maximum generated power for $\gamma_c=10^{-5}\frac{q}{h}$, while tuning the ratio $r=\gamma/\gamma_c$. We note that for high values of $r$ or for low values of $r$, the conductance of the system is determined by $min(\gamma_c,~\gamma)$. For $r\geq 10$, the conductance of the system is mainly determined by $\gamma_c$  and hence increasing $\gamma$ further has negligible effect on the heat harvesting properties of the system. However, as $r$ is gradually decreased from $r=10$, the overall conductance of the system decreases, resulting in a decrease in the generated power. The decrease in generation efficiency with decrease in $\gamma$ is a bit complicated to understand. Let us consider the cycle $I_1 \Rightarrow \ket{0,0,0}\rightarrow \ket{1,0,0}\rightarrow \ket{1,1,0}\rightarrow \ket{0,1,1}\rightarrow \ket{0,1,0}\rightarrow \ket{0,0,0}$. This cycle (explained in the main text) transmits an electron from $L$ to $R$ while absorbing a heat packet $U_m$ from $G$. Note that in this cycle, an electron has to undergo interdot tunneling and hence the rate of this cycle depends on both the interdot coupling $\gamma$ and the system-to-reservoir coupling. Next, let us consider the cycle $I_2\Rightarrow \ket{0,0,0}\rightarrow \ket{1,0,0}\rightarrow\ket{1,1,0}\rightarrow\ket{0,1,0}\rightarrow\ket{0,0,0}$. In this cycle, a heat packet $U_m$ is absorbed from the reservoir $G$ without a net transport of electrons between the reservoirs. Hence, such processes lead to a deterioration of the overall efficiency. Note that in this cycle, an electron doesn't undergo inter-dot tunneling and hence the rate of this process is dependent mainly on $\gamma_c$. As the ratio $r$ is decreased keeping $\gamma_c$ constant, the rate of the cycle $I_1$ decreases while $I_2$ remains unaffected, that is a higher fraction of heat energy is lost without any transport of electrons between $L$ and $R$. This leads to a deterioration in the generation efficiency with decrease in $r$. \\
\indent Fig.~\ref{fig:app_4} demonstrates the variation in heat harvesting performance of the system when the values of $\gamma$ and $\gamma_c$ are interchanged. In Fig.~\ref{fig:app_4}, I demonstrate the heat harvesting performance for (i) $\gamma_c=10^{-5}\frac{q}{h}$ and $\gamma=10^{-4}\frac{q}{h}$, resulting in $r=10$ (shown in green) and (ii)  $\gamma_c=10^{-4}\frac{q}{h}$ and $\gamma=10^{-5}\frac{q}{h}$, resulting in $r=0.1$ (shown in brown). Since the overall conductance of the system, for large and small $r$, is determined by the $min(\gamma_c,~\gamma)$, which is equal to $10^{-5}\frac{q}{h}$ in both the cases, there is negligible variation in the generated power between the two cases. The generation efficiency, however,  decreases drastically for case (ii) when the values of $\gamma$ and $\gamma_c$ are interchanged. This is because a higher value of $\gamma_c$ in case (ii) drastically increases the rate of the cycles $I_2$ (as discussed above), leading to more heat packets being wasted without any net transmission of electrons between $L$ and $R$. This causes a deterioration in the generation efficiency.
 \bibliography{apssamp}
\end{document}